\documentclass[amsmath,amssymb,nofootinbib,superscriptaddress]{revtex4-1}

\usepackage[compact]{titlesec}         
\titlespacing{\subsection}{10pt}{10pt}{10pt}
\titlespacing{\section}{10pt}{10pt}{10pt}

\usepackage[allcolors=blue,colorlinks=true]{hyperref}
\usepackage{amsmath}
\usepackage{tikz-cd}
\usepackage{esint}
\usepackage[symbol]{footmisc}
\numberwithin{equation}{section}

\makeatletter
\renewcommand{\p@subsection}{}
\renewcommand{\p@subsubsection}{}
\makeatother

\newcommand{\cH}{\mathcal{H}}
\newcommand{\cG}{\mathcal{G}}

\newcommand{\ket}[1]{\left|#1\right\rangle}

\renewcommand*{\thefootnote}{\fnsymbol{footnote}}

\begin{document}

\title{\phantom{}\vspace{80pt}\\
\Huge  Quasinormal Modes \\ from \\ Penrose Limits
\vspace{20pt}}

\DefineFNsymbols*{mysymbols}[math]{\approx}
\setfnsymbol{mysymbols}
\author{\Large Kwinten Fransen\footnotemark{\color{blue}$^{\small\approx}$} \\ \phantom{} \\ \large Department of Physics, University of California, Santa Barbara, CA 93106, USA}

\begin{abstract}
	\vspace{20pt}
	We use Penrose limits to approximate quasinormal modes with large real frequencies. The Penrose limit associates a plane wave to a region of spacetime near a null geodesic. This plane wave can be argued to geometrically realize the geometrical optics approximation. Therefore, when applied to the bound null orbits around black holes, the Penrose limit can be used to study quasinormal modes. For instance, this Penrose limit point of view makes manifest the symmetry that emerges in the geometrical optics approximation of quasinormal modes in terms of isometries of the resulting plane wave spacetime.  We apply the procedure to warped $\rm{AdS}_3$, Schwarzschild black holes and Kerr black holes. In the former we show explicitly how the symmetry algebra contracts to that of the limiting plane wave while in the latter we find the expected agreement with numerically computed quasinormal modes for large (real) frequencies.
\end{abstract}

\begin{minipage}{\linewidth}
\maketitle
\end{minipage}
\footnotetext{{\color{blue}kfransen@ucsb.edu}}

\vspace{10pt}

\setcounter{tocdepth}{1}
\begin{minipage}{\linewidth}
\tableofcontents
\end{minipage}
\clearpage

\renewcommand*{\thefootnote}{\arabic{footnote}}
\setcounter{footnote}{0}

\section{Introduction}

The theoretical defining feature of a black hole is its event horizon. Yet, arguably, it is the similarly remarkable existence of bound null geodesics \cite{darwin1959gravity} that leaves the more characteristic imprint on black hole observations. Specifically, such null geodesics play an important role in observations of black holes electromagnetically through their imprint on the black hole image, as observed by the Event Horizon Telescope (EHT) \cite{ames1968optical, Cunha:2018acu, EventHorizonTelescope:2019dse, Gralla:2019xty, EventHorizonTelescope:2022wkp}, and gravitationally through their relation to the black hole ringdown, as observed by gravitational wave detectors \cite{goebel1972comments,Cardoso:2016rao, LIGOScientific:2016aoc}. The latter is based on a  well-known relation between quasinormal modes (QNMs) in the geometrical optics approximation and bound lightlike geodesics \cite{Ferrari:1984zz,Mashhoon:1985cya,Cardoso:2008bp,Dolan:2010wr, Yang:2012he}. These geometrical optics quasinormal modes are essentially determined by the wave equation ``near'' the associated null geodesics or more precisely by a ``near-photon ring'' region in phase space \cite{Hadar:2022xag}. A null Fermi normal coordinate system attached to the geodesic should thus be well-suited to describe quasinormal modes in the geometrical optics approximation. On the other hand, the leading order spacetime that emerges in null Fermi normal coordinates is a plane wave spacetime \cite{Blau:2006ar}. That is to say, the ``near-photon ring'' physics is described by what is known as the Penrose limit of the spacetime \cite{penrose1976any,blau2011plane}
\begin{equation}\label{eqn:planewave}
ds^2 = 2 du dv + A_{ij}(u) x^i x^j du^2+dx_i dx^i \, .
\end{equation}
Here, the latin indices $i$, $j$ run over the transverse directions of the null Fermi-Walker coordinates and $A_{ij}(u) = R_{\mu i \nu j}u^{\mu}u^{\nu}$ is the tidal tensor evaluated on the null geodesic with tangent $u^{\mu}$. This generally depends on an affine coordinate $u$ along the geodesic and it controls the geodesic deviation properties such as the Lyapunov exponent that are in turn reflected in the associated quasinormal modes in the geometrical optics approximation. \\
The plane wave \eqref{eqn:planewave} is, rather literally, the harmonic oscillator of spacetimes. The (null) geodesic equation, in terms of an affine parameter $\tau$, is given in light-cone gauge by 
\begin{equation}\label{eqn:planewavegeodesic}
	\ddot{x}^{i}(\tau) = p_v^2 A^{i}{}_{j}\left(p_v \tau\right) x^{j}(\tau) \, , \qquad u = p_v \tau \, , \qquad p_v v = -\frac{1}{2} x^{i} \dot{x}_i +p_v v_0 \, .
\end{equation}
On the other hand, the wave equation can be reduced to the quantum harmonic oscillator by a Fourier transform with respect to $v$, whose associated momentum $p_v$ will play the role of the inverse Planck constant
\begin{equation} \label{eqn:quantumoscillator}
\frac{i}{p_v} \partial_u \Phi =\left(  \frac{1}{2p_v^2} \delta^{ij}\partial_i \partial_j  + \frac{1}{2} A_{ij}(u) x^i x^j \right) \Phi  \, .
\end{equation}
Conversely, one could have started from the harmonic oscillator and recovered the spacetime \eqref{eqn:planewave} as its Eisenhart-Duval lift \cite{eisenhart1928dynamical,Duval:1990hj}. Consequently, a generic plane wave spacetime has a Heisenberg isometry algebra, with the covariantly constant $\partial_v$ acting as a central element. This algebra, the wave equation \eqref{eqn:quantumoscillator} and everything else about the plane wave \eqref{eqn:planewave} can be explicitly understood in terms of solutions of the geodesic motion \eqref{eqn:planewavegeodesic}. The geometrical properties of the plane wave \eqref{eqn:planewave} are in this sense, like the harmonic oscillator \eqref{eqn:quantumoscillator}, semi-classically exact \cite{Diaz:2004rv}. In the context of the Penrose limit, this means it captures exactly the information that is present in the geometrical optics approximation up to and including the leading amplitude evolution. To summarize, the relation between the geometrical optics approximation, to that order, and the Penrose limit is as follows: \\

\begin{figure}[h]\label{fig:summary}
\begin{tikzcd}
	\phantom{test}\text{General spacetime} \phantom{t} (M,g) \phantom{test} \arrow[r, "\text{Penrose limit}{}_{\gamma}"] \arrow[d," "]& \phantom{test}\text{Plane wave spacetime} \phantom{t} (M_{\gamma},g_{\gamma})\phantom{test} \arrow[d," "] \\
		\phantom{test}\text{Wave equation on $(M,g)$}\phantom{test} \arrow[r, "\text{Geometric optics}{}_{\gamma}"'] & \phantom{test}\text{Wave equation on $(M_{\gamma},g_{\gamma})$}\phantom{test}
\end{tikzcd}
\caption{Geometrical optics approximation, including the leading amplitude evolution, to the wave equation based on the null geodesic $\gamma$ on a spacetime $M$ with metric $g$ is the exact wave equation on the Penrose limit spacetime $M_{\gamma}$ with metric $g_{\gamma}$.}
\end{figure}

The purpose of this paper is to revisit the geometrical optics approximation on black hole spacetimes from the Penrose limit perspective. Specifically, we apply it to quasinormal modes (QNMs) in the geometrical optics approximation. \\

Quasinormal modes in the geometrical optics approximation are well-known \cite{Yang:2012he}. Yet, we believe its reformulation in terms of Penrose limit spacetimes has several advantages. In general, it translates problems from transport equations along a worldline to field theory problems on a plane wave spacetime. As a result, it might be used to bring the wide-array of field theory techniques that have been developed for and applied to plane wave spacetimes to a setting of great astrophysical and observational interest. For quasinormal modes in particular, powerful techniques are potentially available in relation to the fact that, often, the isometry algebra of the plane wave further enhances. For Schwarzschild black holes in particular,the associated plane wave is a (Cahen-Wallach) symmetric space \cite{cahen1970lorentzian}. As a more specific application, and providing the original motivation of this work, the Penrose limit plane wave provides a geometrical point of view on symmetries and properties that emerge in the geometrical optics approximation \cite{Hadar:2022xag,Kapec:2022dvc}. \\

The structure of this paper is as follows. First, in Section \ref{sec:warped}, we discuss in detail the relation between the exact solution to the quasinormal problem, its geometrical optics approximation and the Penrose limit for the toy model of self-dual warped $\rm{AdS}_3$. Subsequently, in Section \ref{sec:schwarzschild} and Section \ref{sec:kerr} respectively, we use the Penrose limit to describe quasinormal modes in the geometrical optics approximation for Schwarzschild and Kerr black holes. We conclude in Section \ref{sec:conclusion} by discussing several research directions for which we believe the geometrical reformulation of the geometrical optics limit will be valuable.

\section{Warped $\rm{AdS}_3$}\label{sec:warped}
The quasinormal mode problem of the three dimensional BTZ black hole \cite{Banados:1992wn} can be solved exactly \cite{Ichinose:1994rg,Birmingham:2001dt}. Therefore, this could be a good starting point to illustrate and explore the main idea. However, a complication arises as there are strictly speaking no circular null geodesics, although one can be argued to exist at the conformal boundary \cite{Kapec:2022dvc}. To avoid this subtlety, we instead follow \cite{Kapec:2022dvc} and use self-dual warped  $\rm{AdS}_3$ as a first toy example. This spacetime arises as a near-horizon region of a near-extreme warped $\rm{AdS}_3$ black hole. Moreover, it is a hypersurface of constant polar angle of the near-horizon region of a near-extremal Kerr black hole \cite{Moussa:2003fc,Bouchareb:2007yx,Bredberg:2009pv}. As a result, the following analysis will also serve as a useful reference later when we discuss the Kerr black hole, in Section \ref{sec:kerr}. \\

The metric of self-dual warped $\rm{AdS}_3$ can be written as \cite{Bredberg:2009pv}
\begin{equation}\label{eqn:metricwarped}
	ds^2 = \ell^2 \left(-r \left(r+4\pi T_H \right) dt^2 + \frac{dr^2}{r \left(r+4\pi T_H \right)} + \Lambda^2 \left(d \phi + \left(r+2\pi T_H \right)dt\right)^2\right) \, .
\end{equation}
Here, $T_H$ is the Hawking temperature while $\Lambda$ is the ``warp factor'' which becomes unity for (ordinary)  $\rm{AdS}_3$. Write the four-velocity of a null geodesic with impact parameter $b = \frac{L}{E}$, the ratio of the energy associated to $\partial_t$ and angular momentum with respect to $\partial_{\phi}$, as
\begin{equation}\label{eqn:warpedfour}
	u_{\mu} dx^{\mu}= - dt + b \,  d\phi + S_r'(r) \, dr \, .
\end{equation}
Then, in the background \eqref{eqn:metricwarped}, it should satisfy
\begin{equation}\label{eqn:warpedgeodesic}
	r(r+4\pi T_H)\left(S_r'(r)\right)^2 = \frac{\left(1+b\left(r+2\pi T_H \right)\right)^2}{r(r+4 \pi T_H)}-\frac{b^2}{\Lambda^2} \, ,
\end{equation}
or, phrased differently
\begin{equation}
	\left(\ell^2 u^r\right)^2 + V(r) = 0 \, , \qquad {\rm with} \qquad  u^r = \frac{1}{\ell^2} r \left(r+4\pi T_H \right) S_r'(r) \, ,
\end{equation}
where the potential is given by
\begin{equation}
	V(r)  = \frac{b^2}{\Lambda^2} r \left(r+4\pi T_H \right)  - \left(1+b\left(r+2\pi T_H \right)\right)^2 \, .
\end{equation}
 $u^{\mu}$ can be complemented to a parallel propagated null frame along the geodesic $u^{\mu}$ generates, by, say, $n^{\mu}$ and $e^{(1)}{}_{\mu}$ satisfying
\begin{equation}
	u^{\mu}n_{\mu} = e^{(1)}{}^{\mu} e^{(1)}{}_{\mu}  = 1 \, , \qquad 	u^{\mu}u_{\mu} = 	n^{\mu}n_{\mu}  = n^{\mu}e^{(1)}{}_{\mu} = u^{\mu}e^{(1)}{}_{\mu} = 0\, .
\end{equation} 
Using such a frame, the Penrose limit can be straightforwardly obtained \cite{Blau:2003dz}. For brevity, and because this is the main case of interest, we immediately restrict to the circular orbits with $V'(r)=V(r)=0$, which occur when
\begin{equation}
	r^{(\pm)}_{0} = 2 \pi T_H \left(\pm \frac{1}{\sqrt{1-\frac{1}{\Lambda^2}}} -1 \right) \, , \qquad b^{(\pm)}_0 = \mp \frac{1}{2 \pi T_H \sqrt{1-\frac{1}{\Lambda^2}}} \, .
\end{equation}
We will typically restrict to orbits outside of the horizon $r^{(+)}_{0}$, even though both play an interesting role \cite{Kapec:2022dvc}. The presence of additional bound photon orbits can spoil the naive relation between null geodesics and quasinormal modes \cite{Khanna:2016yow}. In the spacetime \eqref{eqn:metricwarped}, it can be argued and explicitly shown that this is not the case. However, also more generally, quasinormal modes in the geometrical optics approximation can be characterized by bound null geodesics even if not by a single geodesic. This should be analogous to techniques developed for quantum mechanical applications in the semi-classical limit \cite{gutzwiller1970energy,berry1976closed,richens1982quantisation}. It could be interesting to apply such techniques to the spectrum of quasinormal modes, possibly from the perspective of Penrose limits but, as it is not needed for the black holes of most interests to us, we shall not pursue this here. \\

For the null frame on the circular trajectory $r^{(+)}_{0}$, we shall find it useful to rescale the affine time and use
\begin{equation}
\bar{u}^{\mu} \partial_{\mu} = 2 \pi T_H \ell^2 u^{\mu} \partial_{\mu}   = -\frac{1}{2 \pi T_H} \left( \partial_t + \Omega \partial_{\phi} \right) \, , \qquad \Omega = -2 \pi T_H \sqrt{1-\frac{1}{\Lambda^2}} \, , 
\end{equation}
instead of $u^{\mu} \partial_{\mu} $. To complement this to a full frame, and for future purposes, it is useful to express this in terms of the isometry generators
 \begin{equation}\label{eqn:warpedisometryexplicit}
\begin{aligned}
\hat{L}_0{}^{\mu}\partial_{\mu} &= -\frac{1}{2 \pi T_H} \partial_t \, , \qquad \qquad  W_0{}^{\mu}\partial_{\mu} = \frac{\Omega}{2\pi T_H} \partial_{\phi} \, , \\
\hat{L}_-{}^{\mu}\partial_{\mu} &= \frac{e^{-2\pi t T_H}}{\sqrt{2 r\left(r+4\pi T_H\right)}}  \left(-\frac{r+2\pi T_H}{2\pi T_H}\partial_t - r\left(r+4\pi T_H\right) \partial_r +2\pi T_H \partial_{\phi}\right) \, , \\
\hat{L}_+{}^{\mu}\partial_{\mu}   &= \frac{e^{2\pi t T_H}}{\sqrt{2 r\left(r+4\pi T_H\right)}}  \left(-\frac{r+2\pi T_H}{2\pi T_H}\partial_t + r\left(r+4\pi T_H\right) \partial_r +2\pi T_H \partial_{\phi}\right) \, .
\end{aligned} 
\end{equation}
These generate the group $\rm{SL}(2,\mathbb{R}) \times U(1)$ 
\begin{equation}\label{eqn:warpedalgebra}
\left\lbrack \hat{L}_0, \hat{L}_{\pm} \right\rbrack = \mp \hat{L}_{\pm}  \, , \qquad 	\left\lbrack \hat{L}_+, \hat{L}_{-} \right\rbrack =  \hat{L}_{0} \, ,\qquad \left\lbrack W_0, \hat{L}_{0,\pm} \right\rbrack = 0 \, .
\end{equation}
The (rescaled) four-velocity on the circular orbit can then be written as
\begin{equation}
	\bar{u}^{\mu} \partial_{\mu} = \hat{L}_0{}^{\mu}\partial_{\mu} - W_0{}^{\mu}\partial_{\mu} \, .
\end{equation}
The geodesic thus preserves the particular subgroup of $\rm{SL}(2,\mathbb{R}) \times U(1)$  generated by this combination. Consider now
\begin{equation}\label{eqn:warpedframet0}
	\bar{e}^{(1)}{}^{\mu}\partial_{\mu} = \frac{1}{\sqrt{2} \ell}\left(\hat{L}_+{}^{\mu}  - \hat{L}_-{}^{\mu}\right)\partial_{\mu} \, , \qquad \bar{n}_0{}^{\mu} \partial_{\mu} =  \frac{1-\Lambda^2}{2 \ell^2}\left(\hat{L}_0{}^{\mu} + \frac{1+\Lambda^2}{1-\Lambda^2} W_0{}^{\mu}\right)\partial_{\mu} \, .
\end{equation}
In coordinates, on the null geodesic at the time $t=0$
\begin{equation}
	\bar{e}^{(1)}{}^{\mu}\partial_{\mu} = \frac{2 \pi T_H}{\ell \Lambda \sqrt{1-1/\Lambda^2}} \partial_r \, , \qquad \bar{n}_0{}^{\mu}\partial_{\mu} = \frac{\Lambda^2-1}{4 \pi T_H \ell^2} \left(\partial_t + \Omega \frac{\Lambda^2+1}{\Lambda^2-1} \partial_{\phi}\right) \, .
\end{equation}
Together with $\bar{u}^{\mu}$, these form a null frame at this time. They can straightforwardly be made into a parallel null frame for the entire evolution. However, by symmetry and as we have checked explicitly, \eqref{eqn:warpedframet0} are sufficient to derive the Penrose limit
 \begin{equation}\label{eqn:warpedpenrose}
ds^2 = 2du dv +dx^2 + A x^2 du^2 \, ,
\end{equation}
with
\begin{equation}\label{eqn:warpedA}
A = \left(-R_{\mu \nu \alpha \beta} \bar{u}^{\mu} e^{(1)}{}^{\nu} \bar{u}^{\alpha} e^{(1)}{}^{\beta}\right)_{\gamma} =  1\, .
\end{equation}
Here $u$ is interpreted as the affine parameter (associated to $\bar{u}^{\mu}$) along the null geodesic, $x$ is the transverse coordinate associated to the direction $e^{(1)}{}^{\beta}$ and $v$ is constant along the wave front. $A$ is the ``frequency matrix'' that governs the geodesic deviations away from the background bound null geodesic. We have chosen $\bar{u}^{\mu} \partial_{\mu}$ to normalize \eqref{eqn:warpedA}, but anticipating a conversion back to the ordinary time $t$, one can suggestively write this as
\begin{equation}
	A = 1 =\frac{\gamma_L^2 }{4 \pi^2 T_H^2} \, ,
\end{equation} 
in terms of the Lyapunov exponent
\begin{equation}
	\gamma_L^2  = 4 \pi^2 T_H^2 \, .
\end{equation}
Intrinsically, \eqref{eqn:warpedpenrose} has lost all information about the orbit and the input on the relation between the coordinates used in \eqref{eqn:warpedpenrose} and their origin in the full spacetime is crucial. Phrased differently, one should always view \eqref{eqn:warpedpenrose} to exist within a matched asymptotic expansion with the ambient spacetime \eqref{eqn:metricwarped}. This will specifically be needed to impose boundary conditions. Next, we will show that a massless scalar field on the background \eqref{eqn:metricwarped}, in the geometrical optics approximation near the circular null geodesic, should precisely correspond to solutions of the wave equation  on \eqref{eqn:warpedpenrose}. To do so, we will first solve the wave equation on \eqref{eqn:metricwarped}, then restrict to the quasinormal modes, which in the geometrical optics approximation adhere to the underlying assumption of being associated with the photon ring, and finally relate them to solutions of the wave equation on \eqref{eqn:warpedpenrose}. Along the way, we show how the symmetry algebra \eqref{eqn:warpedalgebra} contracts to the isometry algebra of \eqref{eqn:warpedpenrose}. \\

Consider a mode of definite frequency and angular momentum, diagonalizing $\hat{L}_0$ and $W_0$
\begin{equation}\label{eqn:wavefunctionwarped}
	\Phi = e^{-i \omega t + i m \phi}\psi(r) \, .
\end{equation}
The full massless scalar wave equation is then given by
\begin{equation}\label{eqn:warpedhypergeometric}
	 r(r+4\pi T_H) \psi''(r) + 2(r+2\pi T_H) \psi'(r)+ \left(\frac{\left(\omega + m (r+2\pi T_H)\right)^2}{r \left(r+4\pi T_H \right)} - \frac{m^2}{\Lambda^2} \right) \psi(r) = 0 \, .
\end{equation}
The remaining radial ordinary linear differential equation has regular singular points at both horizons, $-4 \pi T_H$ and $0$, with local behaviors around these singularities
\begin{equation}
	\psi \sim r^{\pm \frac{i}{2}\left(m+\frac{\omega}{2 \pi T_H}\right)}\left(1+O\left(r\right) \right) \, , \qquad \psi \sim \left(r+4\pi T_H \right)^{\pm \frac{i}{2} \left(m-\frac{\omega}{2 \pi T_H}\right)} \left(1+O\left(r+4\pi T_H\right)\right) \, ,
\end{equation}
as well as a third asymptotic regular singularity, at $r \to \infty$
\begin{equation}
 \psi \sim \left(\frac{1}{r} \right)^{\frac{1}{2} \pm \sqrt{\frac{1}{4}-m^2+\frac{m^2}{\Lambda^2}}} \left(1+O\left(\frac{1}{r}\right) \right)\, .
\end{equation}
These singularities may be summarized in terms of the exponents
\begin{equation}\label{eqn:warpedmonodromy}
\alpha_0^{(\pm)} = \pm \frac{i}{2}\left(m+\hat{\omega}\right) \, ,\qquad \alpha_{1}^{(\pm)} = \pm \frac{i}{2} \left(m-\hat{\omega}\right) \, ,\qquad \alpha_{\infty}^{(\pm)} = \pm \sqrt{\frac{1}{4}-m^2+\frac{m^2}{\Lambda^2}} = \pm \beta \, ,
\end{equation}
where, for ease of comparison, we follow \cite{Kapec:2022dvc} in introducing
\begin{equation}
\hat{\omega} = \frac{\omega}{2 \pi T_H} \, , \qquad	\beta = i m \sqrt{\left(1-\frac{1}{\Lambda^2}\right) - \frac{1}{4 m^2}}\, .
\end{equation}
 The solutions to \eqref{eqn:warpedhypergeometric} can thus be expressed in terms of the hypergeometric functions with the appropriate branch cut structure imposed by these local solutions. We wish to impose boundary conditions at $r\to 0$ and $r \to \infty$, starting with the former a convenient choice of independent solutions is
\begin{equation}
	\begin{aligned}
	\psi^{\rm in} &= r^{\alpha_0^{(-)}} \left(\frac{r}{4\pi T_H } +1 \right)^{\alpha_1^{(-)}} {}_2F_1\left(\frac{1}{2}+\alpha_{\infty}^{(+)}-\alpha_{1}^{(+)}-\alpha_{0}^{(+)}, \frac{1}{2}-\alpha_{\infty}^{(+)}-\alpha_{1}^{(+)}-\alpha_{0}^{(+)};1-2 \alpha_{0}^{(+)}; -\frac{r}{4 \pi T_H} \right)\, , \\
	\psi^{\rm out} &=  r^{\alpha_0^{(+)}} \left(\frac{r}{4\pi T_H } +1 \right)^{\alpha_1^{(-)}} {}_2F_1\left(\frac{1}{2}+\alpha_{\infty}^{(+)}-\alpha_{1}^{(+)}-\alpha_{0}^{(-)}, \frac{1}{2}-\alpha_{\infty}^{(+)}-\alpha_{1}^{(+)}-\alpha_{0}^{(-)};1-2 \alpha_{0}^{(-)}; -\frac{r}{4 \pi T_H} \right)\, ,
	\end{aligned}
\end{equation} 
or in terms of the physical parameters
\begin{equation}
\begin{aligned}
\psi^{\rm in} &= r^{- \frac{i}{2}\left(m+\hat{\omega}\right)} \left(\frac{r}{4\pi T_H } +1 \right)^{- \frac{i}{2}\left(m-\hat{\omega}\right)} {}_2F_1\left( \frac{1}{2}+\beta-im, \frac{1}{2}-\beta-im;1-i\left(m+\hat{\omega} \right) ; -\frac{r}{4 \pi T_H} \right)\, , \\
\psi^{\rm out} &=  r^{ \frac{i}{2}\left(m+\hat{\omega}\right)} \left(\frac{r}{4\pi T_H } +1 \right)^{- \frac{i}{2}\left(m-\hat{\omega}\right)} {}_2F_1\left( \frac{1}{2}+\beta+i\hat{\omega}, \frac{1}{2}-\beta+i\omega;1+i\left(m+\hat{\omega}\right) ; -\frac{r}{4 \pi T_H} \right)\, ,
\end{aligned}
\end{equation}
as derived also in \cite{Kapec:2022dvc}. \\

The quasinormal mode condition imposes that $\psi^{\rm in}$ has definite monodromy around $r \to \infty$ in addition to $r \to 0$\footnote{This is more subtle in asymptotically flat space where the asymptotic singular point is irregular \cite{Motl:2003cd}.}. Using the explicit connection matrix \cite{Castro:2013lba} or, equivalently, the contiguous relations for the hypergeometric functions \cite{NIST:DLMF}, one finds the quantization condition \cite{Kapec:2022dvc}
\begin{equation}
	n = -\frac{1}{2}\pm\beta + i \hat{\omega} \, ,
\end{equation}
resulting in the quasinormal mode frequencies
\begin{equation}
	\hat{\omega}_{mn (\pm)} = \pm  m \hat{\Omega} \sqrt{1-\frac{1}{4 m^2}\left(1-\frac{1}{\Lambda^2}\right)^{-1}} - i (n+\frac{1}{2}) \, , \qquad \hat{\Omega} = \frac{\Omega}{2 \pi T_H}\, .
\end{equation}
The corresponding wavefunctions are
\begin{equation}\label{eqn:warpedqnmwave}
	\Phi_{mn (\pm)} \propto e^{-i \omega_{mn (\pm)} t + i m \phi} \left(\frac{r}{4 \pi T_H}\right)^{\left(-im \pm \beta-\frac{1}{2}\right)/2} \left(\frac{r}{4\pi T_H } +1 \right)^{\left(im \pm \beta-\frac{1}{2}\right)/2} P_n^{(\frac{1}{2}\mp \beta)}\left(m;i z\right)\, ,
\end{equation}
with
\begin{equation}\label{eqn:meixnerpollaczek}
	P_n^{(\lambda)}(m;\psi) = \frac{(2 \lambda)_n}{n!} e^{i n \psi} {}_2F_1\left(-n,\lambda+im,2\lambda,1-e^{-2 i \psi} \right) \, ,
\end{equation}
and
\begin{equation}\label{eqn:defxwarped}
	z = \frac{1}{2} \ln\left(1+\frac{4\pi T_H}{r}\right) \, , \qquad r = \frac{4 \pi T_H}{\left(e^{2z}-1\right)}  \, .
\end{equation}
This is consistent with the fact that the quasinormal modes associated to the near-horizon region of near-extremal Kerr are related to Meixner-Pollaczek polynomials \cite{Chen:2017ofv}\footnote{A useful application of this is the identification of their generating function in the  zero-damped quasinormal piece of the  time-domain  Green's function on a near-extremal Kerr black hole \cite{Gralla:2016sxp,Compere:2017hsi,Compere:2019wfw}.} as well as more generally the known connection between these polynomials and representations of $\rm{SL}(2,\mathbb{R})$ \cite{koornwinder1988group}. It is expected that spin can be included with an essentially identical functional form, although the interpretation of $\Phi_{mn (\pm)} $ would generally be more complicated. For some properties of the Meixner-Pollaczek polynomials see \cite{chihara2011introduction} as well as \cite{koelink1998convolutions, bender1987continuous,koornwinder1989meixner} for specifically how they naturally arise in the context of representation theory. The first three Meixner-Pollaczek polynomials are given by
\begin{equation}\label{eqn:MPexamples}
\begin{aligned}
 P_0^{(\lambda)}\left(m;i z\right) &= 1 \, , \\
 P_1^{(\lambda)}\left(m;i z\right) &= 2im \sinh(z)+ 2 \lambda \cosh{(z)} \, , \\
 P_2^{(\lambda)}\left(m;i z\right) &= -2 m^2 \sinh ^2(z)+i (2 \lambda +1) m \sinh (2
 z)+\lambda  (\lambda +(\lambda +1) \cosh (2 z)) \, .
\end{aligned}
\end{equation}
It is somewhat curious that these are orthogonal polynomials\footnote{They are orthogonal with respect to a particular measure, see for instance \cite{koornwinder1988group}.} in $m$, not $z$. \\

To take the geometrical optics limit, let $m \sim \beta \to \infty$ and $r \to r_0^{(+)}$. Specifically let
\begin{equation}
r = r_0^{(+)} + 2 \pi T_H \frac{\delta r}{\ell \Lambda \sqrt{1-1/\Lambda^2}} \, , \qquad \delta r \sim \frac{1}{\sqrt{m}} \to 0 \, , 
\end{equation}
where the latter scaling can, for instance, be anticipated from the relevant scalings in the Penrose limit. First, consider the fundamental $n=0$ mode in this limit
\begin{equation}\label{eqn:warpedqnmwave0limit}
\Phi_{m0 (+)} \propto \exp{\left(-i  m \left(\Omega t + \phi\right)-\frac{\lambda_L t}{2}- i m \hat{\Omega} \frac{\delta r^2}{2 \ell^2}\right)} \, .
\end{equation}
Based on the results of \cite{Kapec:2022dvc,Chen:2022fpl}, one can expect that this same limit for $n>0$ will relate the Meixner-Pollaczek polynomials to Hermite polynomials and one readily conjectures
\begin{equation}\label{eqn:MPtoHermiteconj}
	 P_n^{(\frac{1}{2}- \beta)}\left(m;i z\right) \sim \frac{1}{n!}\left( im \hat{\Omega} \right)^{n/2} H_n\left(\sqrt{i m  \hat{\Omega}} \frac{\delta r}{\ell}\right) \, ,
\end{equation}
which can be checked against many examples \eqref{eqn:MPexamples} or, to show it in general, one can first note
\begin{equation}
\frac{1}{2}-	\beta \sim im \hat{\Omega} \, , \qquad iz \sim  i \hat{\Omega} \frac{\Lambda}{\ell} \, , 
\end{equation}
and subsequently appeal to known asymptotic relations in the mathematical literature on orthogonal polynomials \cite{ferreira2003asymptotic}. We thus have, in the geometrical optics approximation
\begin{equation}\label{eqn:warpedqnmwaveindirecht}
\Phi_{mn (+)} \propto e^{-i  m \left(\Omega t + \phi\right)-\lambda_L t(\frac{1}{2}+n)- i m \hat{\Omega} \frac{\delta r^2}{2 \ell^2} } H_n\left(\sqrt{i m  \hat{\Omega}} \frac{\delta r}{\ell}\right) \, .
\end{equation}
However, instead of going into the details of this derivation, it is more illuminating to point out that the relation \eqref{eqn:MPtoHermiteconj} is a (known) direct result of the underlying algebraic structure \cite{koornwinder1988group,Das:2002ij}. In \eqref{eqn:warpedalgebra}, we have already given the full algebra associated to the $\rm{SL}(2,\mathbb{R}) \times U(1)$ isometry group as generated by $\left\lbrace \hat{L}_0,  \hat{L}_+, \hat{L}_{-}, W_0{}\right\rbrace$ and noted that the wavefunctions \eqref{eqn:wavefunctionwarped} diagonalize the Cartan subalgebra  $\left\lbrace \hat{L}_0, W_0{}\right\rbrace$, with eigenvalues $i \hat{\omega}$ and $i m \hat{\Omega}$ respectively. In addition, $ \hat{L}_{\pm} $ act as raising and lowering operators on the QNM wavefunctions \cite{Kapec:2022dvc,Chen:2017ofv}. However, when acting on these QNM wavefunctions in the geometrical optics limit with $\beta \sim im \sim i\hat{\omega} \to \infty$ the algebra contracts to a Heisenberg algebra \cite{Das:2002ij}. Explicitly, recalling the realization of the algebra \eqref{eqn:warpedisometryexplicit}, the action of
\begin{equation}
	  \bar{u}^{\mu}\partial_{\mu} = \hat{L}_0 - W_0  \, ,
\end{equation} 
 remains finite as
 \begin{equation}
 	\hat{\omega}_{mn (+)} -  m \hat{\Omega} \approx - i \left(n+ \frac{1}{2} \right) \, ,
 \end{equation}
while 
 \begin{equation}
   \bar{n}^{\mu}\partial_{\mu} = \frac{1-\Lambda^2}{2 \ell^2}\left(\hat{L}_0{} + \frac{1+\Lambda^2}{1-\Lambda^2} W_0{}\right) \sim \frac{i m}{\ell^2} \hat{\Omega}  \, ,
 \end{equation}
 scales as $m$. On the other hand, to leading order and again expanding around the photon ring by $r = r_0^{(+)} + 2 \pi T_H \frac{\delta r}{\ell \Lambda \sqrt{1-1/\Lambda^2}}$ 
 \begin{equation}\label{eqn:warpedisometryexpanded}
 \begin{aligned}
 \hat{L}_-{}^{\mu}\partial_{\mu} &\approx -\frac{\ell}{\sqrt{2}} e^{-2\pi t T_H}  \left( \partial_{\delta r} -  \delta r \bar{n}^{\mu}\partial_{\mu} \right) \, , \\
 \hat{L}_+{}^{\mu}\partial_{\mu} &\approx \frac{\ell}{\sqrt{2}} e^{2\pi t T_H}  \left( \partial_{\delta r} +  \delta r \bar{n}^{\mu}\partial_{\mu} \right) \, , \\
 \end{aligned} 
 \end{equation}
 which scales as $\sqrt{m}$. Based on these observations, define the rescaled operators
 \begin{equation}
 	\cH = -\hat{L}_0 + W_0 \, , \qquad \hat{h} =  \frac{\epsilon^2}{\ell^2} \frac{1-\Lambda^2}{2}\left(\hat{L}_0{} + \frac{1+\Lambda^2}{1-\Lambda^2} W_0{}\right) \, , \qquad  a^{\dagger} = \frac{\epsilon}{\ell} \hat{L}_- \, , \qquad  a = \frac{\epsilon}{\ell} \hat{L}_+ \, .
 \end{equation}
 These generate the following contracted algebra on geometric optics wavefunctions when $\epsilon^2 \sim 1/m \to 0$
\begin{equation}\label{eqn:warpedalgebracontracted}
\left\lbrack \cH, a \right\rbrack = -a  \, , \qquad \left\lbrack \cH, a^{\dagger} \right\rbrack = a^{\dagger}  \, , \qquad 	\left\lbrack a, a^{\dagger} \right\rbrack = \hat{h} \, ,\qquad \left\lbrack \hat{h}, a \right\rbrack = \left\lbrack \hat{h}, a^{\dagger} \right\rbrack = \left\lbrack \hat{h}, \cH \right\rbrack = 0 \, .
\end{equation}
The essential part of the contraction is that the rescaled expressions for $\left\lbrack a, a^{\dagger} \right\rbrack$, only allows $\bar{n}^{\mu}\partial_{\mu}$ part to make a nonzero contribution as $\epsilon \to 0$, while, similarly, in the other commutators only the constant action of $\bar{n}^{\mu}\partial_{\mu}$ survives. The algebra thus reduces to the centrally extended Heisenberg algebra, with an additional outer action of $\cH$. It can also be described as a central extension of the two-dimensional Poincar\'e algebra. The contracted algebra remains well-defined in the ${\rm AdS}_3$ limit $\Lambda \to 1$. This contraction was also effectively found in \cite{Kapec:2022dvc,Chen:2022fpl}. \\

 To give another perspective on the algebraic considerations in the Penrose limit more generally, its formulation in terms of null Fermi normal is particularly useful \cite{Blau:2006ar}. Consider first a choice of Riemann normal coordinates at a point. There is a residual Lorentz symmetry  that remains unfixed. In addition, such a normal coordinate expansion associates a scale to the asymptotic coordinate ranges. Specifically, Riemann normal coordinates expand in distances small as compared to a characteristic local curvature scale $R$. The residual coordinate changes should respect this implicit underlying scale. For instance, if a particular direction would cover a range that is parametrically larger, the superficial symmetry of the asymptotic metric could be modified as one would have to consider the effect of (potentially amplified) subleading corrections upon mixing of the coordinates. This is what happens when going to Fermi normal coordinates. In principle, both coordinate systems have the same leading form but the time direction now covers ranges on the order of the curvature scale $R$, parametrically larger than the spatial directions. That is, as opposed to $x \sim ct$, one would for general finite $x$ and $t$ have $x \sim \frac{v}{c} ct$ for a dimensionless ratio $\frac{v}{c} \ll 1$ on account of $\frac{x}{R} \ll 1$ while $ct \sim R$. A choice of Fermi normal coordinates based on the associated timelike geodesic then effectively has a residual ``Galilean'' (asymptotic) symmetry group. It is in that sense an In\"on\"u-Wigner contraction of the Riemann normal coordinates \cite{inonu1953contraction}\footnote{Conversely, one could say In\"on\"u-Wigner use a type of Fermi normal coordinates associated to an abelian subgroup on a group manifold for their construction.}. The Penrose limit or null Fermi normal coordinates could similarly be associated with a more general type of contraction \cite{saletan1961contraction, levy1967deformation}. \\

So far the analysis has been qualitatively identical to that in \cite{Kapec:2022dvc}. However, let us now reinterpret these result by returning to the Penrose limit spacetime \eqref{eqn:warpedpenrose}. First, this has a Heisenberg isometry algebra corresponding exactly to \eqref{eqn:warpedalgebracontracted}. That is, the isometries $\lbrace \partial_u, \partial_v, \hat{L}_{\pm}^{(\rm p)} \rbrace$ with 
\begin{equation}
	\hat{L}^{(\rm p)}_{\pm} =  \frac{e^{\pm u}}{\sqrt{2}} \left(\partial_x  \mp x \partial_v \right) \, ,  \qquad  
\end{equation}
or
 \begin{equation}
\cH^{(\rm p)} = -\partial_u\, , \qquad \hat{h}^{(\rm p)} =  \frac{\epsilon^2}{\ell^2} \partial_v \, , \qquad  a_{(\rm p)} = \frac{\epsilon}{\ell} \hat{L}^{(\rm p)}_+ \, , \qquad  a_{(\rm p)}^{\dagger} = \frac{\epsilon}{\ell} \hat{L}_-^{(\rm p)} \, ,
\end{equation}
satisfy
\begin{equation}\label{eqn:warpedalgebraplane}
\left\lbrack \cH^{(\rm p)} , a_{(\rm p)} \right\rbrack = -a_{(\rm p)}  \, , \qquad \left\lbrack \cH^{(\rm p)} , a^{\dagger}_{(\rm p)} \right\rbrack = a^{\dagger}_{(\rm p)}  \, , \qquad 	\left\lbrack a_{(\rm p)}, a^{\dagger}_{(\rm p)} \right\rbrack = \hat{h}^{(\rm p)}  \, ,\qquad \left\lbrack \hat{h}^{(\rm p)} , a_{(\rm p)} \right\rbrack = \left\lbrack \hat{h}^{(\rm p)} , a^{\dagger}_{(\rm p)} \right\rbrack = \left\lbrack \hat{h}^{(\rm p)} , \cH^{(\rm p)}  \right\rbrack = 0 \, .
\end{equation}
Therefore, entirely consistent with the interpretation of the Penrose limit coordinates based on $\bar{u}^{\mu}\partial_{\mu}$, $\bar{n}^{\mu}\partial_{\mu}$ and $\bar{e}^{(1)}{}^{\mu}\partial_{\mu}$, we can indeed identify \eqref{eqn:warpedalgebraplane} with the isometry algebra of the original spacetime \eqref{eqn:metricwarped} acting on geometric optics quasinormal modes
\begin{equation}\label{eqn:warpedalgebracontractedpp}
\cH^{(\rm p)} = \cH \, , \qquad \hat{h}^{(\rm p)} =   \hat{h}  \, , \qquad  \hat{L}^{(\rm p)}_{\pm} = \hat{L}_{\pm}  \, .
\end{equation}
Consider now a massless scalar field on the Penrose limit plane wave
\begin{equation}\label{eqn:wavefunctionwarpedplane}
\tilde{\Phi} = e^{ i p_v v + i p_u u} \tilde{\psi}(x) \, .
\end{equation}
It should satisfy
\begin{equation}
	\frac{1}{2 p_v^2} \tilde{\psi}'' =   \left(\frac{p_u}{p_v} - A \frac{x^2}{2}\right) \tilde{\psi} \, .
\end{equation}
This can be solved in terms of parabolic cilinder functions. However, imposing additionally outgoing boundary conditions, that is radiation is leaking out from the photon ring, one finds the quantization condition
\begin{equation}
	p_u =  -  \frac{i}{2} \sqrt{A} \left(2n+1\right) \, ,
\end{equation}
with associated wavefunctions
\begin{equation}
 \tilde{\Phi}_{n} \propto e^{ i p_v v + i p_u u-i \sqrt{A} p_v \frac{x^2}{2}} H_n\left( \left(i \sqrt{A} p_v \right)^{1/2} x \right) \, .
\end{equation}
Alternatively, one could have of course taken an algebraic approach based on \eqref{eqn:warpedalgebraplane}. \\

Let us finally relate the coordinates back to those of the original spacetime \eqref{eqn:metricwarped}, specifically, $v$ is associated o the (rescaled) action \eqref{eqn:warpedfour} and $u$ to the proper time \cite{Patricot:2003dh, Blau:2004yi}
\begin{equation}
	 v = 2 \pi T_H \ell^2 \left(- t - \frac{1}{2 \pi T_H \sqrt{1-\frac{1}{\Lambda^2}}} \phi \right)\,  , \qquad u = -2 \pi t_H t  \, .
\end{equation}
Thus the wavefunction can be written as
\begin{equation}
	\tilde{\Phi} = \exp{\left( -i p_v  2 \pi T_H \ell^2 \left(t+ \frac{1}{2 \pi T_H \sqrt{1-\frac{1}{\Lambda^2}}} \phi \right) - i 2 \pi T_H p_u t \right)} \tilde{\psi}(x)  \, .
\end{equation}
Periodicity in $\phi$ implies the quantization condition
\begin{equation}
	2 \pi T_H \ell^2 p_v = m \Omega = -2 \pi T_H m \sqrt{1-\frac{1}{\Lambda^2}} \, ,
\end{equation}
such that
\begin{equation}\label{eqn:warpedplanewavefinalcomparison}
\tilde{\Phi}_{m n} \propto e^{ -i m \Omega t + m \phi - \lambda_L (n+\frac{1}{2}) t -i m \hat{\Omega} \frac{x^2}{2 \ell^2}} H_n\left( \sqrt{i m \hat{\Omega}} \frac{x}{\ell} \right)  \, .
\end{equation}
One thus indeed finds that \eqref{eqn:warpedplanewavefinalcomparison} corresponds exactly to the geometrical optics approximation \eqref{eqn:warpedqnmwaveindirecht}
\begin{equation}
	\tilde{\Phi}_{m n} \propto
	\left(\Phi_{mn (+)}\right)_{|\rm geometrical \, \,  optics} \, ,
\end{equation}
when appropriately identifying $\delta r = x$. \\

In summary, the Penrose limit reproduces the physics of the ``near-photon ring'' region described in \cite{Hadar:2022xag, Kapec:2022dvc}. This is conceptually our main result. It yields an insightful geometric perspective on \cite{Hadar:2022xag, Kapec:2022dvc} and, in the remainder, we shall give more astrophysical examples and further argue that this reformulation is useful.

\section{The Schwarzschild black hole}\label{sec:schwarzschild}

Having established the basic principle in the last section, consider now the example of a static black hole described by the Schwarzschild metric
\begin{equation}
	ds^2 = g_{\mu \nu}dx^{\mu}dx^{\nu} =-\left(1-\frac{2M}{r}\right)dt^2 + \left(1-\frac{2M}{r}\right)^{-1}dr^2+r^2 d\Omega^2 \, .
\end{equation}
The massless scalar wave equation
\begin{equation}
	\frac{1}{\sqrt{-g}}\partial_{\mu} \left(\sqrt{-g} g^{\mu \nu} \partial_{\nu} \Psi\right) \, , \qquad g = {\rm det}\left(g_{\mu\nu}\right) \, ,
\end{equation} 
can be reduced to its radial component 
\begin{equation}
\left(\partial^2_{r_*}+\omega^2 - \left(1-\frac{2M}{r}\right)\left(\frac{l(l+1)}{r^2} + \frac{2M}{r}\right) \right) \psi(r) = 0 \, , \quad \frac{d r_*}{d r} = \left(1-\frac{2 M}{r}\right)^{-1} \, ,
\end{equation}
or alternatively
\begin{equation}
\left( \frac{1}{r^4} \partial^2_{r_c}+\omega^2 - (1-\frac{2M}{r})\frac{l(l+1)}{r^2}\right) \frac{\psi(r)}{r} = 0  \, , \qquad \frac{d r_c}{d r} = \frac{1}{r (r-2M)} \, ,
\end{equation}
by the separation into definite frequency partial waves 
\begin{equation}\label{eqn:schwarzschildseparation}
\Psi = \frac{\psi(r)}{r}e^{- i\omega t}Y_{\ell m}(\theta,\phi) \, .
\end{equation}
The geometrical optics limit, for the reduced one-dimensional problem, amounts to the WKB approximation
\begin{equation}
\left(\frac{1}{r^2}\frac{d S_r}{dr_c}\right)^2 = 1 - \left(1-\frac{2M}{r}\right)\frac{\ell (\ell+1)}{\omega^2 r^2} \, , 
\end{equation}
whose solutions are related to the solutions of the radial Hamilton-Jacobi equation for null geodesics with tangent $u^{\mu}$
\begin{equation}\label{eqn:HJschwarzschild}
	g^{\mu \nu}u_{\mu}u_{\nu} =  (1-\frac{2 M}{r})^{-1}\left(-1+\frac{1}{r^4}\left(\partial_{r_c}S_r\right)^2\right)+\frac{b^2}{r^2} = 0 \, , \qquad \frac{\psi(r)}{r} = A(r)e^{i \omega S_r(r)} \, .
\end{equation} 
Indeed, let
\begin{equation}\label{eqn:lframestatic}
	u_{\mu}dx^{\mu} =  d\left(- t + b \phi + S_r(r) \right)  \, ,
\end{equation}
and note
\begin{equation}
\begin{aligned}
	g_{\mu \nu} dx^{\mu} dx^{\nu} = \left(1-\frac{2 M}{r}\right)\left(-dt^2+ r^4 dr_c^2\right)+r^2 \left(d\theta^2+\sin^2{\theta} d\phi^2\right) \, ,
\end{aligned}
\end{equation}
to recover \eqref{eqn:HJschwarzschild} as the Hamilton-Jacobi equation when identifying $b^2 = \frac{\ell(\ell+1)}{\omega^2}$. This relation, as well as its extension to the amplitude $A$, is well-known \cite{Yang:2012he}. However, we will approach the problem here again from the Penrose limit point of view. An important difference with the previous case arises as we now have two transverse directions and, in the full wave equation, one is solvable exactly. Using the Penrose limit, not only the radial but also the angular direction will only be captured approximately. Although this is perhaps undesirable here, in more complicated cases it allow for a uniform treatment of the different transverse directions and it avoids relying on a separation of variables. Therefore, although taking the Penrose limit would be more complicated, we see no unsurmountable obstruction to perform the same procedure when separation of variables is not possible, in contrast to the WKB approximation. In fact, even for Kerr, which is still separable, having a combined, unseparated (in the usual sense) approach for the radial and angular directions is useful, as we will discuss in Section \ref{sec:kerr}.  \\

The Penrose limits of the Schwarzschild black hole are well-known  \cite{Blau:2003dz}, but let us nevertheless briefly construct it. Choose a null frame aligned with $u^{\mu}$, as in \eqref{eqn:HJschwarzschild}. Specifically, we use 
\begin{equation}\label{eqn:nframestatic}
	n_{\mu}dx^{\mu} = \frac{1}{2} (1-\frac{2M}{r}) d\left( t + b \phi + S_r \right)  \, ,
\end{equation}
and
\begin{equation}\label{eqn:eframestatic}
e^{(1)}_{\mu}dx^{\mu} = \left(\frac{b}{r} dr - (r-2M)\partial_r S_r d\phi \right) \, , \qquad e^{(2)}_{\mu}dx^{\mu} = r d\theta \, ,
\end{equation}
such that
\begin{equation}
u_{\mu} u^{\mu} = n_{\mu} n^{\mu} = 0 \, , \quad n_{\mu} u^{\mu} = 1  \, , \quad	e^{(i)}{}_{\mu}e^{(j)}{}^{\mu}=\delta^{ij} \, ,
\end{equation}
with $i$ and $j$ running over the transverse directions $1,2$. This is not yet parallel in general, but as in the previous example it suffices for our purposes. The associated Penrose limit plane wave in Brinkmann coordinates is given by
\begin{equation}\label{eqn:ppSchwarzschildgeneral}
ds^2 = 2 du dv + A_{ij} x^i x^j du^2+dx_1^2+dx^2_2 \, .
\end{equation}
Here
\begin{equation}
	A_{ij} = \left(-R_{\mu \nu \alpha \beta} u^{\mu} e^{(i)}{}^{\nu} u^{\alpha} e^{(j)}{}^{\beta}\right)_{\gamma}  \, ,
\end{equation}
evaluated on the geodesic $\gamma$ with affine coordinate $u$, is given explicitly by
\begin{equation}\label{eqn:schwarzschildAexpl}
 A_{11} = -A_{22} = \frac{3 b^2 M}{r(u)^5} \, ,\qquad A_{12}=A_{21}=  0 \, ,
\end{equation}
where it is implied that $r=r(u)$. However, on the photon ring
\begin{equation}
r=r_0 = 3M \, , \qquad \frac{3}{2} \frac{b^2M}{r_0^5} = \frac{1}{6 M^2} \, , \qquad b =3 \sqrt{3}  M \, ,
\end{equation}
and the frame \eqref{eqn:lframestatic}-\eqref{eqn:eframestatic} simplifies to
\begin{equation}\label{eqn:staticframephotonring}
	u_{\mu}dx^{\mu} = -dt +  3\sqrt{3} M d\phi \, , \qquad 	n_{\mu}dx^{\mu} = \frac{1}{6}\left(dt + 3\sqrt{3} M d\phi \right) \, , \qquad e^{(1)}_{\mu}dx^{\mu}  = \sqrt{3} dr \, , \qquad e^{(2)}_{\mu}dx^{\mu} = r_0 d \theta \, .
\end{equation}
In particular, the transverse directions are aligned with $r$ and $\phi$ respectively. Moreover, from the solution to the geodesic equation in the original Schwarzschild coordinates
 \begin{equation}\label{eqn:staticphotonringoriginalcoordinates}
t_{\gamma}(u) = 3 u \, , \qquad r_{\gamma}(u) = r_0 \, , \qquad \theta_{\gamma}(u) = \frac{\pi}{2} \, , \qquad \phi_{\gamma}(u) = \phi_0+ \frac{u}{\sqrt{3} M} \, .
\end{equation}
 From the frame and \eqref{eqn:schwarzschildAexpl}, one deduces the familar fact that perturbations in the radial ($x_1$) direction are unstable with the characteristic Lyapunov timescale $\lambda_L$. On the other hand, perturbations in the angular ($x_2$) direction will cause a precession around the original orbit with frequency $\omega_{\rm prec}$. For this specific spherically symmetric example, the latter will of course simply change the overall orientation of the circular trajectory, as seen from compatibility with the orbital frequency $\omega_{\rm orb}$. Explicitly
 \begin{equation}\label{eqn:schwarzschilddeviationfrequencies}.
 	\lambda_L = \frac{1}{\sqrt{27} M} \, , \qquad  \omega_{\rm prec} = \omega_{\rm orb} =  \frac{1}{\sqrt{27} M} \, .
 \end{equation} 
 All these observations descend directly from the classic results for geodesic deviations. Moreover, it is no coincidence here that  $\lambda_L = \omega_{\rm prec}$, one associated to a stable and the other to an unstable direction, as \eqref{eqn:ppSchwarzschildgeneral} should still satisfy the vacuum Einstein equations. \\
 
\begin{figure}[t!]
	\begin{center}
		\includegraphics[width=0.48\textwidth]{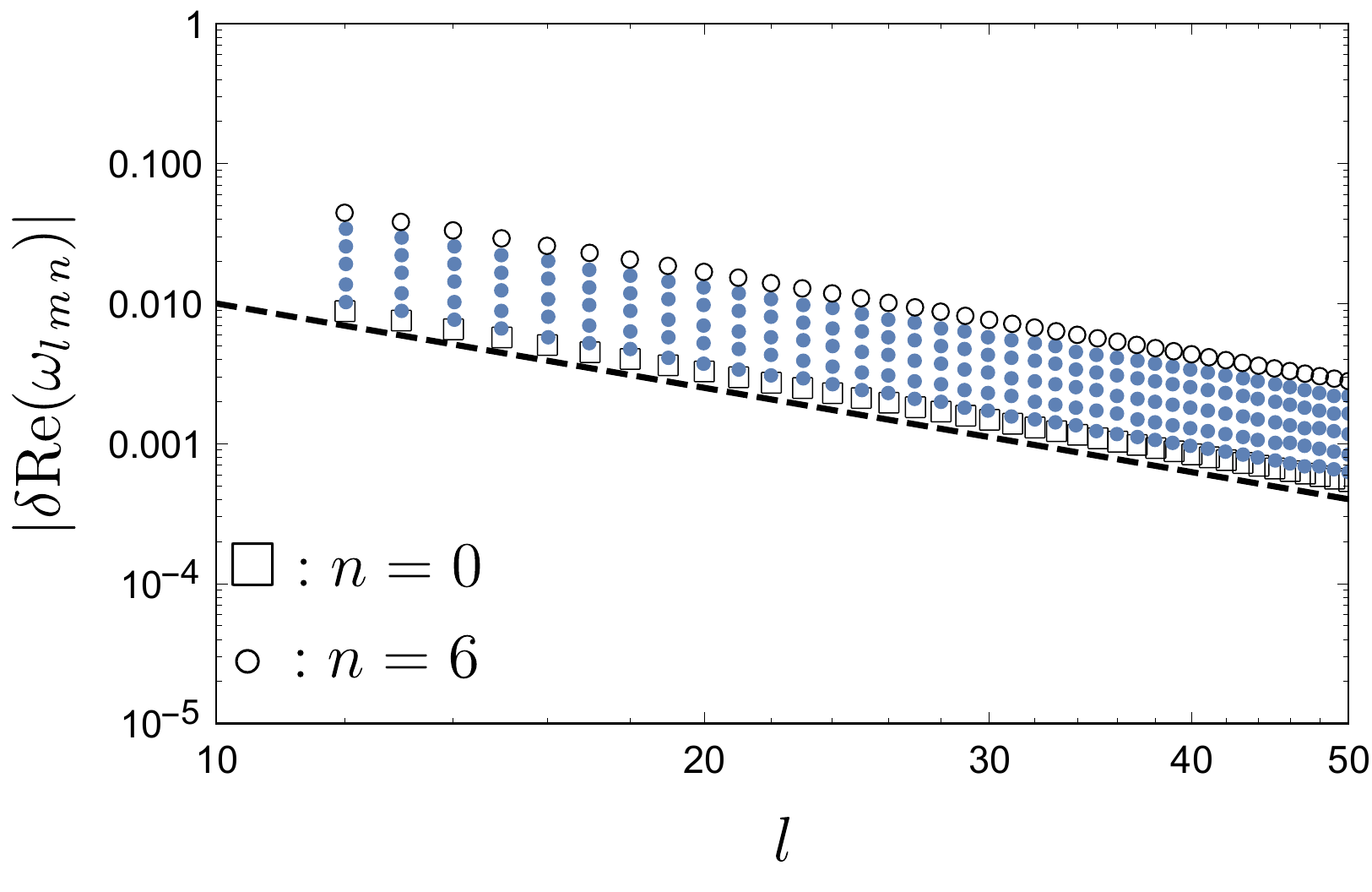} 
		\includegraphics[width=0.48\textwidth]{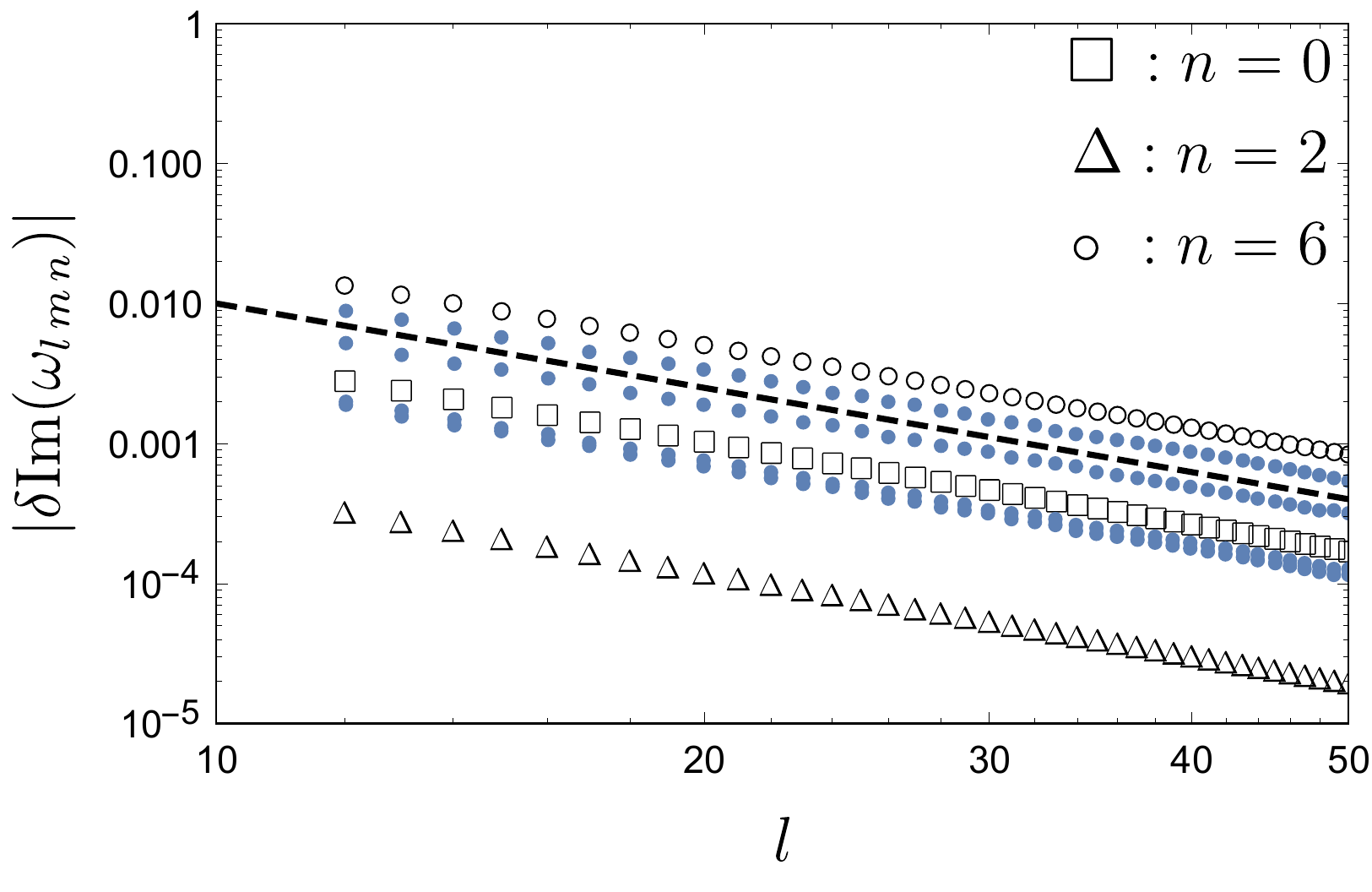}
		\caption{Comparison of the geometrical optics approximation \eqref{eqn:schwarzschildQN2M} to the Schwarzschild quasinormal modes with their numerical values, for overtones $n=0$ (squares) to $n=6$ (circles) as a function of the angular mode number $l$. Specifically, what is shown is the relative difference, which we denote as  $\delta x = \left(x_{\rm approx}-x_{\rm num}\right)/x_{\rm num}$. The convergence with the angular mode number $l$ follows a $l^{-2}$-scaling as indicated by the (black) dashed reference line. This is expected for the real part of the quasinormal mode frequency but better than one should generally expect for the imaginary part.}
		\label{fig:Schwarzschildqnms}
	\end{center}
\end{figure}

 The exact solution to the wave equation on the background \eqref{eqn:ppSchwarzschildgeneral} proceeds as in the previous section. We separate
\begin{equation}\label{eqn:schwarzschildwavefunctionplane}
\tilde{\Phi} = e^{ i p_v v + i p_u u} \tilde{\psi}_1(x_1) \tilde{\psi}_2(x_2) \, ,
\end{equation}
and find
\begin{equation}\label{eqn:schwarzschildwaveabstract}
	\begin{aligned}
	\frac{1}{2 p_v^2} \tilde{\psi}_1''(x_1) + A_{11} \frac{x^2}{2} \tilde{\psi}_1(x_1) &= \left(\frac{p_u}{2 p_v} + K\right) \tilde{\psi}_1(x_1)    \, , \\
	\frac{1}{2 p_v^2} \tilde{\psi}_2''(x_2) + A_{22} \frac{x^2}{2} \tilde{\psi}_2(x_2) &= \left(\frac{p_u}{2 p_v} - K\right) \tilde{\psi}_2(x_2)    \, , \\
	\end{aligned}
\end{equation}
using already that $A_{ij}$ is constant and $A_{12} = A_{21} = 0$. Both equations are again solved in terms of parabolic cilinder functions in general. For the quasinormal modes, we demand an outgoing boundary condition for the unstable direction ($x_1$), and, additionally, we impose a decaying boundary condition for the stable direction ($x_2$). Combined, these yield the quantization condition
\begin{equation}
	p_u = i \sqrt{A_{11}} \left(\frac{1}{2}+n_1\right) -  \sqrt{-A_{22}} \left(\frac{1}{2}+n_2\right)\, .
\end{equation}
 Lastly, $p_v$ is fixed by a quantization condition from the action \eqref{eqn:lframestatic}, or specifically from periodicity of $\phi$ in \eqref{eqn:schwarzschildwavefunctionplane}\footnote{Here, we could have still let $l < 0$, but it would be more natural to include this possibility in a change of sign for $b$. For instance from the identification $b = \frac{\sqrt{l(l+1)}}{\omega} \approx \frac{l+\frac{1}{2}}{\omega}$}
 \begin{equation}
 	p_v = \frac{L}{b} = L \omega_{\rm orb}  =\frac{L}{3 \sqrt{3} M} \, , \qquad  1 \ll L \in \mathbb{N} \, .
 \end{equation}
 We can thus conclude with the  Schwarzschild quasinormal modes in the geometrical optics limit
 \begin{equation}\label{eqn:schwarzschildQNM}
 	\omega_{L n_1 n_2} = L \omega_{\rm orb}+\left(\frac{1}{2}+n_2\right)\omega_{\rm prec} -i \lambda_L \left(\frac{1}{2}+n_1\right) \, ,
 \end{equation}
 as well as their wavefunctions
 \begin{equation}\label{eqn:schwarzschildwavefunctionplanefinal}
 \begin{aligned}
 \tilde{\Phi}_{L n_1 n_2} = & \exp{\left( -i \omega_{L n_1 n_2} t + i L \phi + i 9 L \omega_{\rm orb} \lambda_L \frac{\delta r^2}{2}-  27 L \omega_{\rm orb} \omega_{\rm prec} \frac{M^2 \delta \theta ^2}{2}\right)}  \\ &\times H_{n_1}\left(\sqrt{-i 9 L \omega_{\rm orb} \lambda_L } \delta r \right) H_{n_2}\left(3\sqrt{3 L \omega_{\rm orb} \omega_{\rm prec}}  M \delta \theta\right) \, .
 \end{aligned}
 \end{equation}
 To find these, recall from \eqref{eqn:schwarzschildAexpl} that $A_{11} = -A_{22} = \frac{1}{3 M}$, while also using the relations to the original Schwarzschild coordinates 
  \begin{equation}
\sqrt{3} \delta r =  x_1 \, , \qquad r_0 \delta \theta =  x_2 \, ,  \qquad t(u) = 3 u  \, .
 \end{equation}
In \eqref{eqn:schwarzschildQNM}, we have kept the form of \eqref{eqn:schwarzschildQNM} that best expresses the structure of the geometrical optics quasinormal modes as it will emerge in general complicated examples, but here it can of course be simplified using $ \omega_{\rm orb} = \omega_{\rm prec}$
 \begin{equation}\label{eqn:schwarzschildQN2M}
 \omega_{l n} = \left( l + \frac{1}{2}\right)\omega_{\rm orb} -i \lambda_L \left(\frac{1}{2}+n\right) \, , \qquad l = L + n_2 \, , \qquad n = n_1 \, .
 \end{equation}
 The interpretation of $n_1$ as the usual overtone number comes from its relation to the zeros of the wavefunction in the radial direction while those of $L = m$ and $n_2 = l-m$ are similarly established for the angular directions. This result is compared to the numerical quasinormal modes in Figure \ref{fig:Schwarzschildqnms}. We find \eqref{eqn:schwarzschildwavefunctionplanefinal} matches the ``near-ring'' analysis of \cite{Hadar:2022xag}, where it was obtained from a traditional geometrical optics perspective to approximate the wave equation on the full Schwarzschild spacetime. Instead, we have reformulated it in terms of a Penrose limit.  \\
 
Let us conclude our discussion on the Schwarzschild black hole by a brief remark on the Penrose limit for more general null orbits. These are not related to quasinormal modes but could still be useful in the context of scattering problems or solutions to the wave equation more generally. They will inevitably be more complicated, as due to the time dependence of $r(u)$, one no longer has a symmetric space. Nevertheless, by nature of the approximation, once the geodesics in the plane wave spacetime are understood, or alternatively the harmonic oscillator with the time-dependent frequency matrix $A_{ij}(u)$, all other properties, solutions to the wave equation in  particular, could be derived. Moreover, asymptotically, in the far past of future, one can only be either in asymptotically flat space, in the singularity or at the light ring \cite{Blau:2003dz,Blau:2004yi}. In all of these cases, the plane wave has an increased symmetry that would allow one to practically define asymptotic states to set-up scattering problems in this formulation. \\
 
The reader well-acquainted with group contractions is now invited to move on to the discussion of the Kerr black hole in the next section. However, as it is instructive, we will linger a while longer to discuss the contraction of the associated Legendre polynomials of the full solution to the Hermite polynomials of the Penrose limit or, more generally, representations of the rotation group to those of the Heisenberg algebra. Group contractions of the three dimensional rotation group are one of the original examples of In\"on\"u-Wigner and the result is in fact the two dimensional Euclidean group \cite{inonu1953contraction}. Explicitly, define the rescaled angular momenta 
\begin{equation}\label{eqn:3drotationredef}
L_z = m \hat{L}_z \, , \qquad L_x = \sqrt{\frac{l(l+1)}{m}} \hat{L}_x \, , \qquad L_y = \sqrt{\frac{l(l+1)}{m}}  \hat{L}_y\, ,
\end{equation} 
in terms of the Cartesian coordinates $x$, $y$, $z$. Then as $l \to \infty$
\begin{equation}\label{eqn:rescaledrotations}
[\hat{L}_x, \hat{L}_y]= \frac{m^2 \hat{L}_z}{l(l+1)}  \to 0 \, , \qquad 	[\hat{L}_z, \hat{L}_x]= \frac{1}{m} \hat{L}_y \, , \qquad [\hat{L}_z, \hat{L}_y]= -\frac{1}{m}\hat{L}_x \, .
\end{equation}
In\"on\"u-Wigner observe that representations of the rotation group are no longer faithful representations of the contracted group. However, they suggest restoring this by considering simultaneously a limiting sequence of representations. This has been our perspective from the start. Explicitly, denote the relevant angular part of the partial wave solutions \eqref{eqn:schwarzschildseparation} to the wave equation on the Schwarzschild background as $\ket{m,l}$. By construction, these are (say normalized) eigenstates of the total (rescaled) angular momentum
\begin{equation}\label{eqn:hatquadraticcasimir}
\hat{L}^2 = \frac{m^2}{l(l+1)}\hat{L}_z^2+\frac{1}{m}\left(\hat{L}_x^2+\hat{L}_y^2\right) \, ,
\end{equation}
 and (rescaled) angular momentum in the $z$-direction (associated to $\phi$ rotations) 
\begin{equation}
\hat{L}^2 \ket{m,l} = \ket{m,l} \, , \qquad	\hat{L}_{z} \ket{m,l} = \ket{m,l} \, .
\end{equation}
One finds the familiar result
\begin{equation}\label{eqn:Lpmaction}
\hat{L}_{\pm} \ket{m,l}  = \sqrt{\frac{m}{l(l+1)}} N_{l,m}\ket{m \pm 1,l} \, , \qquad N_{l,m} = \sqrt{l(l+1)-m(m \pm 1) } \, ,
\end{equation}
for the raising and lowering operators
\begin{equation}
\hat{L}_{\pm} = \hat{L}_x \pm i \hat{L}_y \, .
\end{equation}
Therefore, \eqref{eqn:Lpmaction} stays finite as $l\to \infty$ and one indeed recovers a faithful representation of the two-dimensional Euclidean group. For instance, let $\hat{x} = \rho \cos{\phi}$ and $\hat{y} = \rho \sin{\phi}$ for the explicit example of the representation of the contracted algebra on the plane, where $\hat{L}_x$ and  $\hat{L}_y$ generate respectively translations in the $\hat{x}$ and $\hat{y}$ directions. These define the natural variables to separate the Laplacian, the remainder of the quadratic Casimir \eqref{eqn:hatquadraticcasimir} as $l \to \infty$, while still diagonalizing $\hat{L}_z$. The associated radial equation will have solutions in terms of Bessel functions and In\"on\"u-Wigner use this to derive a relation between the associated Legendre polynomials and the relevant Bessel functions \cite{inonu1953contraction, watson1922treatise}. Implicit here is that $m$ is not of the order of $l$, contrary to our limit where $m \sim l$. In fact, the only non-trivial contraction of the three-dimensional rotation group, in the original sense of In\"on\"u-Wigner and the generalization of Saletan, is the one we have just discussed \cite{saletan1961contraction}. Nevertheless, the more singular scaling of \eqref{eqn:3drotationredef} with $l \sim m \to \infty$, also leads to a well-defined algebra \cite{inonu1964group, levy1967deformation}
\begin{equation}\label{eqn:rescaledrotations2}
[\hat{L}_x, \hat{L}_y]= \frac{m^2 \hat{L}_z}{l(l+1)}   \, , \qquad 	[\hat{L}_z, \hat{L}_x]= \frac{1}{m} \hat{L}_y  \to 0 \, , \qquad [\hat{L}_z, \hat{L}_y]= -\frac{1}{m}\hat{L}_x   \to 0 \, ,
\end{equation}
with a finite action on our partial wave representation of the rotation group, as seen from \eqref{eqn:Lpmaction}. This is (part of) the Heisenberg algebra of the plane wave we were looking for. Consider the explicit representation of the raising and lowering operators on functions on the sphere as $m \sim l \to \infty$ with $\hat{\phi} = m \phi$ and $\frac{\pi}{2}-\hat{\theta} \sqrt{\frac{m}{l(l+1)}} = \theta$ (as suggested by \eqref{eqn:3drotationredef})
\begin{equation}\label{eqn:laddercontraction}
	\hat{L}_{\pm} \to \left(-\partial_{\hat{\theta}} \pm  \frac{m^2}{l(l+1)} \hat{\theta} \hat{L}_z\right) \, .
\end{equation}
These are also raising and lowering operators for the Hermite polynomials. On the other hand, the ``ground state'' contracts as follows
\begin{equation}\label{eqn:weightcontraction}
	P^m_m\left(\cos{\left(\frac{\pi}{2}-\frac{\alpha \hat{\theta}}{\sqrt{m}}\right)}\right) \propto \sin^{m}{\left(\frac{\pi}{2}- \frac{\alpha }{\sqrt{m}}\hat{\theta}\right)} \to e^{- \frac{1}{2}\alpha^2 \hat{\theta}^2}\, ,
\end{equation}
and, without spelling out the details, using \eqref{eqn:laddercontraction} and \eqref{eqn:weightcontraction}, one ultimately finds (up to the normalization conventions) \cite{koekoek1996askey}
\begin{equation}
	\lim_{m \to \infty} \frac{m^{-n/2}}{(2m-1)!!} P_{m+n}^{m}\left(\cos{\left(\frac{\pi}{2}-\frac{\alpha \hat{\theta}}{\sqrt{m}}\right)}\right) = \frac{1}{n!} H_n\left( \alpha \hat{\theta}\right) e^{- \frac{1}{2}\alpha^2 \hat{\theta}^2} \, .
\end{equation}

 \section{The Kerr black hole}\label{sec:kerr}
 
 Let us finally turn to the main case of astrophysical interest, the Kerr black hole. The formulation of the geometrical optics approximation in terms of the Penrose limit becomes more distinctive here and will lead us to an approximation that, although similar to previous work in spirit, differs in its implementation. Specifically, although closely mimicking \cite{Yang:2012he} and earlier investigations \cite{Ferrari:1984zz, Cardoso:2008bp, Dolan:2010wr,Dolan:2011fh} in understanding the geometrical optics quasinormal mode frequencies in terms of spherical null geodesics, we derive a formulation of the approximate wave equation for which a remaining nontrivial angular evolution is seamlessly integrated with the approximate restriction of the radial wave profile. Our approach therefore does not ultimately rely on separate WKB approximations for the radial and angular wave equations. We do still make fundamental use of the separability for many practical aspects of the calculation, but it is not a structural part of argument. \\
 
  In Boyer-Lindquist coordinates the Kerr metric is given by \cite{boyer1967maximal}
 \begin{equation}
 \begin{aligned}
 	ds^2 &= - \frac{\Delta}{\Sigma} \left(dt -  a \sin^2{\theta} d \phi \right)^2 + \frac{\Sigma}{\Delta} dr^2 + \Sigma d \theta^2 + \frac{\sin^2\theta}{\Sigma} \left\lbrack \left(r^2 + a^2\right) d \phi - a dt \right\rbrack^2 \, , \\
 	\Delta &= r^2 - 2 M r + a^2 \, , \qquad  \qquad \Sigma = r^2 + a^2 \cos^2{\theta} \, ,
 	\end{aligned}
 \end{equation}
 for mass $M$ and angular momentum $J = a M$. We will find a solution to the null geodesic equation and an associated frame using the Hamilton-Jacobi equation
 \begin{equation}
 g^{\mu \nu} \partial_{\mu} S \partial_{\nu}S = 0 \, . \qquad u^{\mu} =  \partial^{\mu}S \, .
 \end{equation} 
This equation is separable \cite{carter1968hamilton, carter1968global} and (very) well-studied  for a Kerr metric, see for instance \cite{Gralla:2019ceu, Compere:2021bkk} for recent expositions. The general solutions can be written in the (separated) form
 \begin{equation}
 S = S_{\theta}(\theta) + b \phi + S_r(r) - t \, ,
 \end{equation}
 with
 \begin{equation}\label{eqn:kerractions}
 \begin{aligned}
 S_r(r) &= \int^r dr' \, \frac{\sqrt{R(r')}}{\Delta(r')}   \, , \qquad R(r) =  \left((r^2+a^2)-a b\right)^2-\Delta \left((b-a )^2+\eta \right) \, , \\
 S_{\theta}(\theta) &= \int^{\theta} d\theta' \, \sqrt{\Theta(\theta')} \, ,\qquad \Theta(\theta) = \eta - \cos^2 \theta \left(\frac{b^2}{\sin^2 \theta}-a^2 \right) \, .
 \end{aligned}
 \end{equation} 
 Here, $\eta$ is the (specific) Carter constant associated to  the Killing tensor $K_{ab}$
 \begin{equation}\label{eqn:killingtensor}
 \eta = K_{a b}u^a u^b - (b-a)^2 \, , \qquad K_{ab} = -f_{ac}f^c{}_b \, ,
 \end{equation}
 with $f$ the Killing-Yano two-form
 \begin{equation}\label{eqn:KYform}
 f =  a \cos{\theta} \left(dt - a \sin^2\theta d \phi \right)\wedge dr - r \sin\theta \left( a dt -\left(r^2+a^2\right)d\phi \right) \wedge d\theta \, .
 \end{equation}
 Spherical geodesics  with constant $r_0$, which will play the role of the photon ring for the Schwarzschild black hole, satisfy
 \begin{equation}
 R(r_0) = 0 \, , \qquad R'(r_0) = 0 .
 \end{equation}
 This implies
 \begin{equation}\label{eqn:kerrsphericalconstants}
  b= -\frac{r_0^2(r_0-3 M)+a^2(r_0+M)}{a(r_0-M)} \, , \qquad \eta = -\frac{r_0^3 \left(-4a^2M+r_0(r_0-3 M)^2\right)}{a^2 (r_0-M)^2} \, .
 \end{equation}
 In the equatorial case with $\eta = 0$, the co-and counter rotating photon orbits are thus respectively at\footnote{As for the warped $\rm{AdS}_3$, there is an additional solution behind the horizon. We will not consider it here.}
 \begin{equation}
 r_{{\rm ph}+}  = 2+2\cos\left(\frac{2}{3} \arccos \frac{a}{M} +\frac{4 \pi}{3}\right)\, , \qquad r_{{\rm ph}-}  = 2+2\cos\left(\frac{2}{3} \arccos \frac{a}{M}\right) \, .
 \end{equation}
 In general, the polar motion is governed by
 \begin{equation}\label{eqn:kerrpolar}
 (r^2_0 + a^2 \cos^2 {\theta}) \dot{\theta} = \sqrt{\Theta(\theta)} \, ,
 \end{equation}
 where $\dot{\theta} = \frac{d\theta}{du} = u^{\theta}$ . \eqref{eqn:kerrpolar} has turning points at the zeros of $\Theta(\theta)$\footnote{This should not cause confusion with the $\rm AdS$ radius of Section \ref{sec:warped}.}
 \begin{equation}
 	\left(\sin{\theta_\pm}\right)^2 =  \ell^2\left(1 \pm \sqrt{1+ \frac{\eta}{\ell^2 a^2}} \right) \, , \qquad \ell^2 = \frac{\eta + b^2-a^2}{2 a^2} \, .
 \end{equation}
 On the other hand, the azimuthal angle and Boyer-Lindquist time evolve as
 \begin{equation}\label{eqn:kerrphitevolve}
 \begin{aligned}
 \left(r^2_0 + a^2 \cos^2 {\theta}\right) \dot{\phi} &= -\left(a - \frac{b}{\sin^2{\theta}}\right) + \frac{a \left((r_0^2+a^2)-b a\right)}{\Delta}\, , \\
 \left(r^2_0 + a^2 \cos^2 {\theta}\right) \dot{t} &= \frac{r_0^2+a^2}{\Delta}\left((r_0^2+a^2)-b a \right) - a \left(a  \sin^2 \theta - b\right) \, .
 \end{aligned}
 \end{equation}
  To take a Penrose limit, we again construct a null frame based on the null four-velocity, restricted to the spherical orbits
 \begin{equation}\label{eqn:kerraction}
 	u_{\mu}d x^{\mu} = dS = - dt + b d \phi + S'_{\theta}(\theta) d \theta \, .
 \end{equation}
 The construction of parallel frames along geodesics for a Kerr spacetimes has been investigated previously \cite{vandeMeent:2019cam, Kubiznak:2008zs}. We will simply appeal to these results, and follow \cite{Igata:2019pgb} in particular. We should note here that there are other ways to take a Penrose limit, which can and have also been applied for Kerr geodesic, for instance the construction of an explicit adapted coordinates system as in \cite{Papadopoulos:2020qik}. In addition, the Penrose limit of Kerr geodesics was already discussed in \cite{Hollowood:2009qz}\footnote{We would like to thank A. Lupsasca for bringing this reference to our attention.}. Analytic expressions, in all cases, naturally lean heavily on the underlying symmetry of the Kerr black hole \cite{Frolov:2017kze}. According to the general approach described \cite{Kubiznak:2008zs} for null geodesics, the following complement $u^{\mu}$ to a parallel propagated null frame
 \begin{equation}\label{eqn:kerrframe1}
 	e^{(1)}{}^{\mu} = \frac{1}{C}\left( u^{\alpha} h_{\alpha}{}^{\mu} - u (u^{\alpha}\xi_{\alpha})u^{\mu} \right)\, , \qquad 	e^{(2)}{}^{\mu} = \frac{1}{K}\left(u^{\alpha} f_{\alpha}{}^{\mu} \right) \, ,
 \end{equation}
 and
 \begin{equation}\label{eqn:kerrframe2}
 	n^{\mu} = -\frac{1}{C}  e^{(1)}{}^{\alpha} h_{\alpha}{}^{\mu} + \frac{1}{2 C^4} \left( C_{\beta}{}^{\gamma}C_{\gamma \delta} u^{\beta} u^{\delta} + u^2 (\xi_{\alpha} u^{\alpha})^2 C^2 \right) u^{\mu} \, ,
 \end{equation}
 with
 \begin{equation}
 	K^2 = K_{\alpha \beta} u^{\alpha} u^{\beta} \, , \qquad  C^2 = C_{\alpha \beta} u^{\alpha} u^{\beta} \, , \qquad C_{\alpha \beta} = h_{\alpha \gamma} h_{\beta}{}^{\gamma} \, .
 \end{equation}
 Here, $K_{\alpha \beta}$ is the Killing tensor defined earlier in \eqref{eqn:killingtensor}, based on the Killing-Yano two-form $f_{\mu \nu}$ from \eqref{eqn:KYform}, while $h_{\mu \nu}$ is the conformal Killing-Yano two-form  dual to $f_{\mu \nu}$
 \begin{equation}
 \begin{aligned}
 	h &= \frac{1}{2}d\left\lbrack\left(-r^2+a^2\cos^2\theta\right)d t + a\left(r^2-(a^2+r^2)\cos^2\theta \right) d\phi \right\rbrack \\ &= r \left(dt - a \sin^2\theta d \phi\right) \wedge dr + a \cos{\theta}\sin{\theta} \left( a dt - \left(r^2+a^2\right)d\phi \right) \wedge d \theta \, , 
 \end{aligned}
 \end{equation}
 and $\xi^{\mu}$ is the Killing vector that is constructed from $h_{\alpha \beta}$ as
 \begin{equation}
 	\xi^{\mu} = \frac{1}{3} \nabla_{\beta}h^{\beta \mu} \, .
 \end{equation}
 For us, $\xi^{\mu}$ is simply the Killing vector $\partial_t$. The frame given in \eqref{eqn:kerrframe1}-\eqref{eqn:kerrframe2}, based on \cite{Kubiznak:2008zs}, works in principle for general null geodesics but we will only apply it for the spherical trajectories. We then find explicitly
 \begin{equation}
 C_{\beta}{}^{\gamma}C_{\gamma \delta} u^{\beta} u^{\delta}  = -\frac{4 \Delta r_0^2 \left(r_0^2-a^2\cos^2{\theta}\right)}{(r_0-M)^2} \, , \qquad	K^2 = C^2 = \frac{4 \Delta r_0^2}{(r_0-M)^2} \, , \qquad \xi^{\mu}u_{\mu} = -1 \, .
 \end{equation}
  The full explicit expressions for the frame are rather unwieldy, but in the Schwarzschild limit, one essentially recovers the Schwarzschild frame used in the last section. More generally, for the equatorial trajectory, $e^{(1)}{}^{\mu}$ can be qualitatively thought of as representing the radial direction. \\
  
\begin{figure}[t!]
	\begin{center}
			\includegraphics[width=0.48\textwidth]{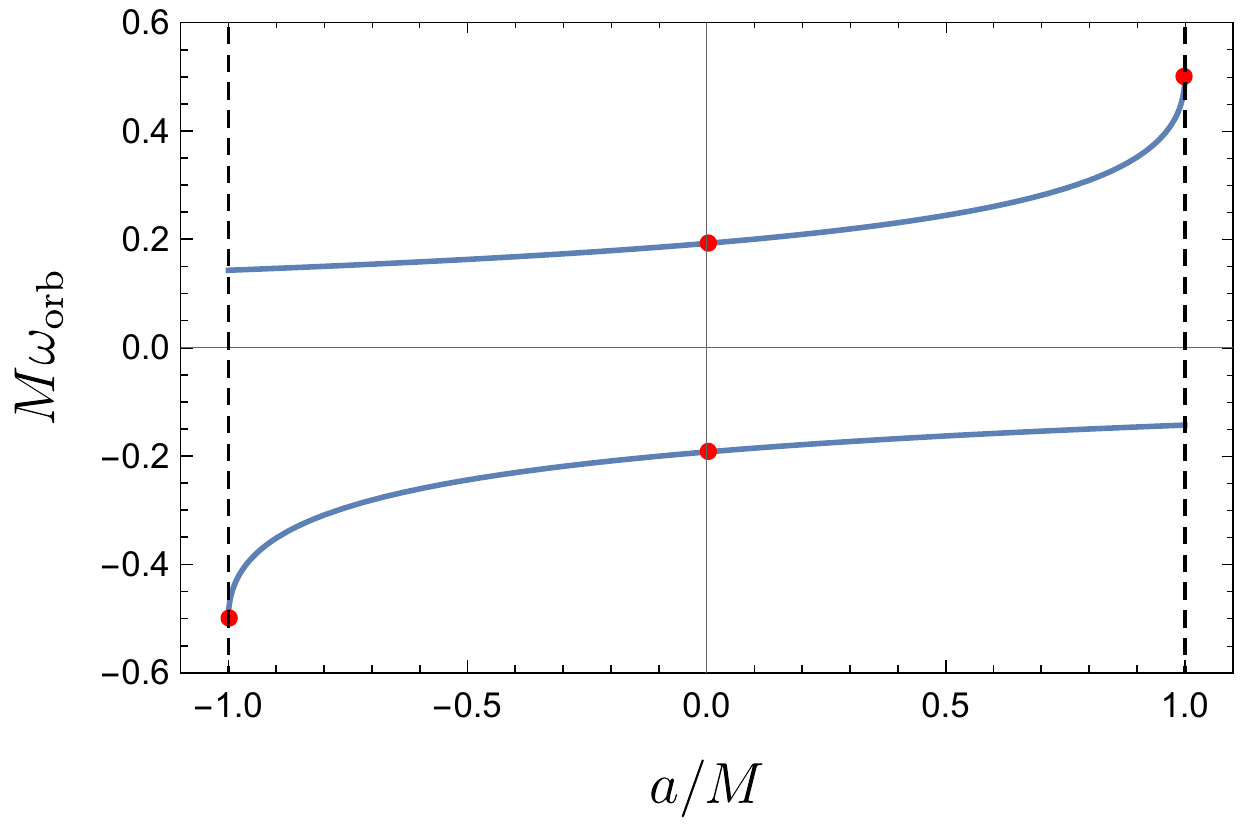} 
			\includegraphics[width=0.48\textwidth]{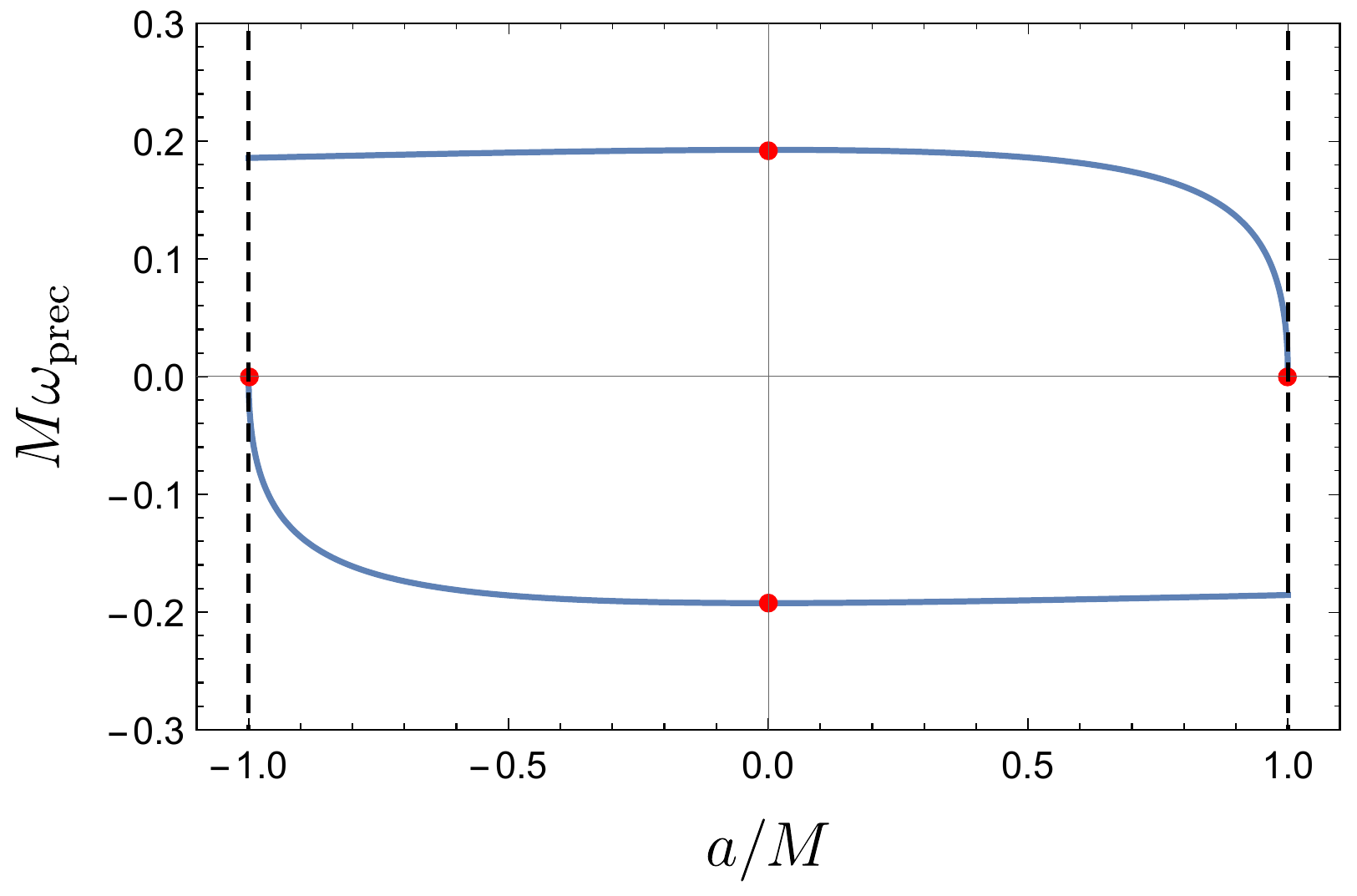}
		\caption{The orbital (left) and precession frequency (right) of equatorial null trajectories \eqref{eqn:kerrEqfreqs}, for $a>0$ the upper (lower) line represents the corotating (counterrotating) orbit. These frequencies fix the associated $l \sim \pm m$ geometrical optics quasinormal modes by \eqref{eqn:kerrEqQNM}. The Lyapunov exponent $\lambda_L$ is related to $\omega_{\rm prec}$ through the Einstein equation. Values of particular interest (red dots) are the Schwarzschild limit $\omega_{\rm orb} = \omega_{\rm prec} = \lambda_L = \frac{1}{3\sqrt{3} M}$ at $a=0$ (full vertical line) and the (corotating) near-horizon near-extremal limit $\omega_{\rm orb} = \frac{1}{2M}$, $\omega_{\rm prec} \, , \lambda_L \to 0$ at $a \to M$ (dashed vertical line).}
		\label{fig:eqfreqfunctions}
	\end{center}
\end{figure}

\begin{figure}[t!]
	\begin{center}
			\includegraphics[width=0.48\textwidth]{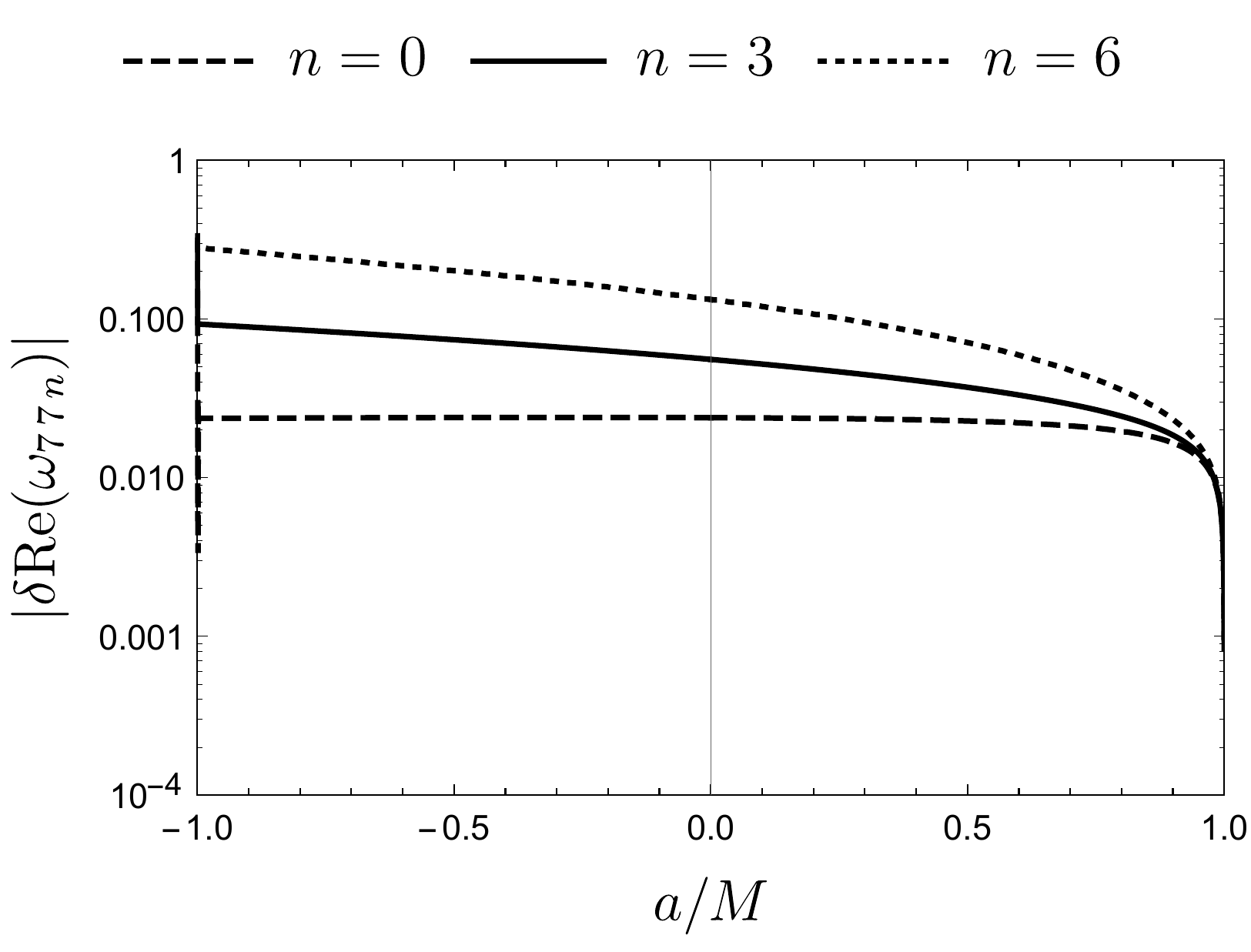} 
			\includegraphics[width=0.48\textwidth]{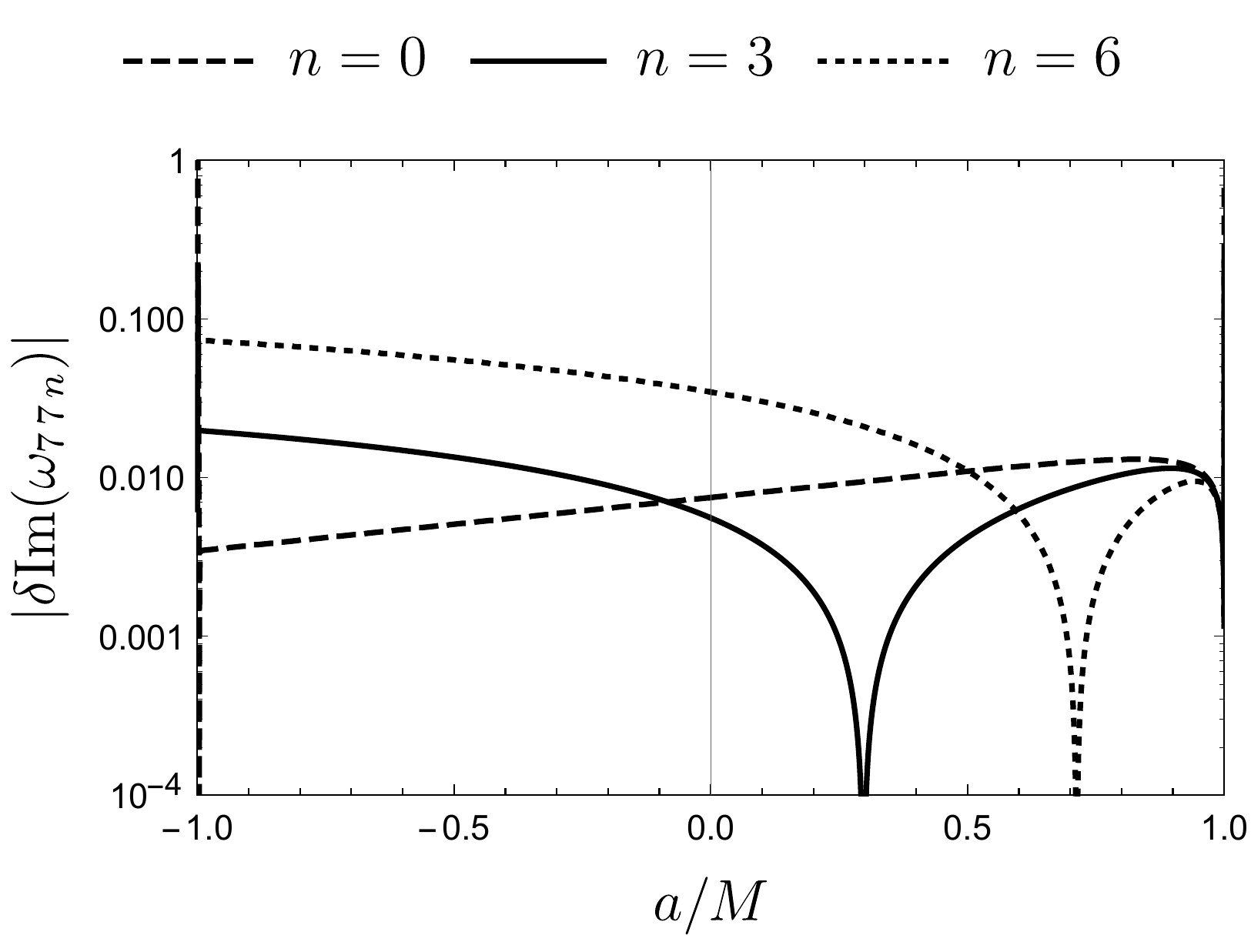}
		\caption{Comparison of the geometrical optics approximation \eqref{eqn:kerrEqQNM}-\eqref{eqn:kerrEqfreqs} to the ``equatorial'' ($l \sim \pm m$) numerically computed Kerr quasinormal mode frequencies for $l=7$ as a function of the black hole angular momentum. For a general quantity $x$, we denote such a relative difference as $\delta x = \left(x_{\rm approx}-x_{\rm num}\right)/x_{\rm num}$. The error is typically larger for higher overtones. The geometrical optics approximation in the corotating extremal limit correctly captures the behavior of the zero-damped quasinormal modes, as observed in \cite{Yang:2012he}. The convergence with $l$ is similar as for the Schwarzschild limit $a=0$ displayed in Figure \ref{fig:Schwarzschildqnms} and is revisited for the general quasinormal modes in Figure \ref{fig:kerrapproxre} and \ref{fig:kerrapproxim}.}
		\label{fig:kerrequatorialapprox}
	\end{center}
\end{figure}

   Using the frame \eqref{eqn:kerrframe1}-\eqref{eqn:kerrframe2}, we find the Penrose limit plane wave
 \begin{equation}\label{eqn:ppkerr}
 \begin{aligned}
 ds^2 &= 2 du dv  +dx_1^2+dx^2_2 + \frac{12 M r^2_0 \Delta du^2}{(r_0-M)^2 \Sigma^5} \\ &\times \left\lbrace  r_0\left\lbrack    5\left(r_0^2-a^2\cos^2{\theta}\right)^2-4r_0^4\right\rbrack \left(x_1^2-x_2^2\right) + 2 a \cos \theta \left\lbrack 5\left(r_0^2-a^2\cos^2{\theta}\right)^2-4 a^4 \cos^4\theta  \right\rbrack x_1 x_2   \right\rbrace \, ,
 \end{aligned}
 \end{equation}
 or
 \begin{equation}\label{eqn:ppkerrgeneral}
 ds^2 = 2 du dv + A_{ij}(u) x^i x^j du^2+dx_1^2+dx^2_2 \, ,
 \end{equation}
 with
 \begin{equation}
 \begin{aligned}
   A_{11} = - A_{22} &= \frac{12 M r^3_0 \Delta}{(r_0-M)^2 \Sigma^5}\left\lbrack 5\left(r_0^2-a^2\cos^2{\theta}\right)^2-4r_0^4\right\rbrack \, , \\
   A_{12} = A_{21} &=   a \cos \theta  \frac{12 M r^2_0 \Delta}{(r_0-M)^2 \Sigma^5} \left\lbrack 5\left(r_0^2-a^2\cos^2{\theta}\right)^2-4 a^4 \cos^4\theta  \right\rbrack  \, .
   \end{aligned}
 \end{equation} 
Contrary to the examples from the previous sections, the general plane wave \eqref{eqn:ppkerr} is time-dependent and not diagonal. The time-dependence comes from $\theta=\theta(u)$ which obeys \eqref{eqn:kerrpolar}. However, for the equatorial orbits, this becomes constant and the problem simplifies. Specifically, it becomes exactly like the Schwarzschild black hole discussed in the previous section. Using the same notation for the separate massless scalar wave equation as in \eqref{eqn:schwarzschildwavefunctionplane}
\begin{equation}\label{eqn:kerrsepeartedwavefunctionplane}
\tilde{\Phi} = e^{ i p_v v + i p_u u} \tilde{\psi}_1(x_1) \tilde{\psi}_2(x_2) \, , \qquad \nabla_{\mu}\nabla^{\mu} \tilde{\Phi} = 0 \, ,
\end{equation}
The wave equation reduces to \eqref{eqn:schwarzschildwaveabstract}
\begin{equation}\label{eqn:kerreqabstract}
\begin{aligned}
\frac{1}{2 p_v^2} \tilde{\psi}_1''(x_1) + A_{11} \frac{x^2}{2} \tilde{\psi}_1(x_1) &= \left(\frac{p_u}{2 p_v} + K\right) \tilde{\psi}_1(x_1)    \, , \\
\frac{1}{2 p_v^2} \tilde{\psi}_2''(x_2) + A_{22} \frac{x^2}{2} \tilde{\psi}_2(x_2) &= \left(\frac{p_u}{2 p_v} - K\right) \tilde{\psi}_2(x_2)    \, . \\
\end{aligned}
\end{equation}
with $K$ a separation constant. The only difference as compared to Schwarzschild is that now
\begin{equation}
	A_{11} = - A_{22} = \frac{12 M \Delta}{r_0^3 (r_0-M)^2} \, .
\end{equation}
Here, $r_0$ should be chosen to be the constant radial coordinate of either equatorial photon ring, $r_{\rm ph \, +}$ or $r_{\rm ph \, -}$. Appealing to the Schwarzschild analysis, we find
 \begin{equation}\label{eqn:kerrEqQNM}
\omega_{L n_1 n_2} = L \omega_{\rm orb}+\left(\frac{1}{2}+n_2\right)\omega_{\rm prec} -i \lambda_L \left(\frac{1}{2}+n_1\right) \, ,
\end{equation}
 with
\begin{equation}\label{eqn:kerrEqfreqs}
\omega_{\rm orb} = \frac{3M-r_0}{a\left(r_0+3M\right)} \, , \qquad  		\omega_{\rm prec}^2 = \lambda_L^2 = \frac{12 M \Delta}{r_0^3 \left(r_0+3M\right)^2} \, .
\end{equation}
These functions are illustrated in Figure \ref{fig:eqfreqfunctions}. The connection with the usual quantum numbers for the quasinormal mode frequency $\omega_{l m n}$ can be made as for Schwarzschild by $l = |L| + n_2$, $m = L$, $n = n_1$. However, some care should be taken in the choice of signs with respect to our previous Schwarzschild analysis, in relation to the boundary conditions on the plane wave background which determine the relative sign between the $\omega_{\rm orb}$ and $\omega_{\rm prec}$ contributions. To be consistent with our previous considerations, this should be positive. With this identification, we can make a comparison with the numerically computed quasinormal mode frequencies with $m \sim l$, as illustrated in Figure \ref{fig:kerrequatorialapprox} for $l=7$. We would like to stress that our notation for $\omega_{\rm orb}$ and $\omega_{\rm prec}$ is generally different from that of \cite{Yang:2012he}.  For equatorial orbits, $\omega_{\rm orb}$ is the leading frequency associated to the ``background'' $\phi$-motion, while $\omega_{\rm prec}$ is a correction associated to the frequency of the $\theta$-motion for a slightly perturbed equatorial geodesic. On the other hand, in the language of \cite{Yang:2012he}, the precession frequency is associated to the azimuthal angle deficit $\Delta \phi_{\rm prec} = \Delta \phi \mp 2 \pi$ of the (co-or counter rotating) $\phi$-motion,  with respect to the traversed angle $\Delta \phi$ in a single $\theta$-period  $T_{\theta} = \frac{2 \pi}{\Omega_{\theta}}$. Therefore, rewriting \eqref{eqn:kerrEqQNM} in the language of \cite{Yang:2012he} yields
\begin{equation}
	\omega_{l m n} = \left( l +\frac{1}{2}\right) \Omega_{\theta}+ m \Omega_{\rm prec} -i \lambda_L \left(\frac{1}{2}+n\right) \, , \qquad \Omega_{\theta} = \omega_{\rm prec} \, , \qquad \Omega_{\rm prec} = \omega_{\rm orb} - \omega_{\rm prec} \, , 
\end{equation} 
 consistent with \cite{Yang:2012he}. \\

Another limit in which one could expect a simplification is the coupled limit where the black hole angular momentum is taken towards extremality $a \to M$ in appropriate coordinates such that it effectively zooms-in on the horizon in a way that resolves the co-rotating equatorial spherical null geodesic from the event horizon \cite{Bardeen:1999px}. As can be seen in Figure \ref{fig:kerrequatorialapprox}, the geometrical optics limit correctly captures the leading behavior in this limit. However, one mainly expects a simplification in the radial direction, or more generally the constant $\theta$-slices, which we have already trivialized by taking the Penrose limit. Let us nevertheless consider the near-horizon near-extremal limit (near-NHEK) for a moment. Based on the discussion of this limit in \cite{Igata:2019pgb}, one could conclude that the Penrose limit is simply flat space. However, that conclusion was based on a (rescaled) frame as constructed by the method of \cite{Kubiznak:2008zs}, which is not guaranteed to work when the conformal Killing-Yano tensor degenerates, which happens in the near-horizon near-extremal limit. We revisit the limit here. We are interested specifically in the near-NHEK limit, which is a scaling of $a \to M$ such that
\begin{equation}
	T = \frac{t}{4 \pi M T_H} \lambda \, , \qquad R = 2 \pi T_H \frac{r-r_+}{M \lambda} \, , \qquad \varphi = \phi - \frac{t}{2M} \, ,
\end{equation}  
remain finite with
\begin{equation}
	\lambda = \sqrt{1-a^2} \, , \qquad r_+ = M\left(1+\lambda\right) \, .
\end{equation}
It is this scaling limit that contains the near-horizon null geodesics at finite $R$ and whose constant $\theta$ slices will correspond to the warped $\rm{AdS}_3$, discussed in Section \ref{sec:warped}, with nonzero temperature. It is given by
\begin{equation}\label{eqn:metricnearNHEK}
\begin{aligned}
ds^2 &= 2M^2 \Gamma(\theta) \left(-R \left(R+4\pi T_H \right) dT^2 + \frac{dR^2}{R \left(R+4\pi T_H \right)}+d\theta^2 + \Lambda^2 \left(d \varphi + \left(R+2\pi T_H \right)dT\right)^2\right) \, , \\
\Gamma &= \frac{1+\cos^2\theta}{2} \, , \qquad \qquad  \Lambda = \frac{2 \sin{\theta}}{1+\cos^2{\theta}} \, .
\end{aligned}
\end{equation}
We find no problem with the frame and recover for the equatorial geodesic the result of Section \ref{sec:warped} up to a rescaling of $u^{\mu}$ and $n^{\mu}$.  The full Penrose limit spacetime is given by 
 \begin{equation}\label{eqn:ppnearNHEK}
\begin{aligned}
ds^2 &= 2 du dv  +dx_1^2+dx^2_2 + \frac{12 R_0 \left(R_0+4 \pi T_H\right) du^2}{M^2 \left(R_0+2\pi T_H\right)^2 \left(1+\cos^2{\theta} \right)^5} \\ &\times \left\lbrace  \left\lbrack    1-10\cos^2{\theta}+5\cos^4{\theta} \right\rbrack \left(x_1^2-x_2^2\right) +  \frac{1}{4}\cos \theta \left\lbrack 3-36\cos2\theta+\cos 4\theta  \right\rbrack x_1 x_2   \right\rbrace \, .
\end{aligned}
\end{equation}
As expected, at least on first inspection, there seems to be no substantial simplification as compared to the plane wave associated to spherical null trajectories for a general Kerr black hole. Specifically, one still has a time-dependent non-diagonal $A_{ij}$. In the usual separation of variables, this is consistent with the nontrivial angular wave equation that persists in this limit. Similarly, the polar motion for a general Kerr black hole at $b=0$ does not seem to simplify the Penrose limit. Therefore, we now turn to the analysis of the general plane wave \eqref{eqn:ppkerr}. \\

To use the Penrose limit plane waves \eqref{eqn:ppkerr} to approximate general non-equatorial quasinormal modes for a rotating black hole, first observe that the quantization condition for $p_v$ already becomes more interesting. We will appeal to the identifications made in \cite{Yang:2012he}, which essentially means imposing a Bohr-Sommerfeld, or more generally an Einstein-Brillouin-Keller (EBK),  quantization condition as argued for on the basis of having a single-valued wavefunction \cite{einstein1917quantensatz,brillouin1926remarques, keller1958corrected}. The EBK quantization condition allows us to express $\eta$ and $b$ in terms of the quantum numbers $l$, $m$ \cite{Yang:2012he}
\begin{equation}\label{eqn:EBK}
p_v \int_{\theta_-}^{\theta_{+}}\, d \theta \sqrt{\Theta} = \left(l-|m| \right)  \pi \,, \qquad  \left(p_v b\right) = m \, . 
\end{equation}
The former equation contains an elliptic integral. It can thus be decomposed in terms of \cite{Gralla:2019ceu}
\begin{equation}
\begin{aligned}
	I_K &= \int_{\theta_-}^{\theta_{+}}t \frac{d \theta}{\sqrt{\Theta}} = \frac{2}{\sqrt{-z_- a^2}} {\rm Ke}\left(\frac{z_+}{z_-}\right) \, , \\	I_{\Pi} &= \int_{\theta_-}^{\theta_{+}} \frac{\csc^2\theta \, d \theta}{\sqrt{\Theta}} = \frac{2}{\sqrt{-z_- a^2}} {\rm \Pi e}\left(z_+ \left| \right. \frac{z_+}{z_-}\right) \,  , \\ 	I_{E} &= \int_{\theta_-}^{\theta_{+}} \frac{\cos^2{\theta} d \theta}{\sqrt{\Theta}} = -\frac{4 z_+}{\sqrt{-z_- a^2}} {\rm Ee}'\left(\frac{z_+}{z_-}\right) \, ,
\end{aligned}
\end{equation}
with ${\rm Ke}$ and $\Pi{\rm e}$ complete elliptic integrals of the first and third kind respectively and ${\rm Ee}'$ related to the complete elliptic integral of the second kind through
\begin{equation}
{\rm Ee}'(\varphi \left| \right. k) = \partial_k {\rm Ee}(\varphi \left|k \right. ) \, .
\end{equation}
We can thus rewrite \eqref{eqn:EBK} as
\begin{equation}
\pi \left(l-|m| \right) = p_v \left(\frac{m^2}{p_v^2}+ \eta \right)I_K -\frac{m^2}{p_v} I_{\Pi} + a^2 p_v I_E \, .
\end{equation}
In the equatorial limit, we recover our previous quantization condition \eqref{eqn:kerrEqfreqs}. On the other hand, in the limiting (polar) case when $b =m = 0$, one finds \cite{Dolan:2010wr}
\begin{equation}
	p_v = l \Omega_{\theta} \, , \qquad  \frac{2 \pi}{\Omega_{\theta}} = 2 \int_{\theta_-}^{\theta_{+}} \frac{dt}{d\theta} d\theta = 4 \sqrt{a^2+\eta} {\rm E e}\left(\frac{ a^2}{a^2+\eta}\right) \, .
\end{equation}
That is, in the limiting cases of azimuthal or polar motion it is respectively the azimuthal and polar frequency that determine the leading real part of the quasinormal mode frequency in the geometrical optics approximation \cite{Dolan:2010wr}. The stable orthogonal direction will determine the first correction to this real frequency, corresponding respectively to a polar or azimuthal precession. More generally, given a desired quasinormal mode associated to $l$, $m$, these mode numbers can be used to fix the geodesic with \eqref{eqn:EBK}. In addition, it will impose a quantization condition on $p_v$, as $b$ and $\eta$ are related for spherical orbits\footnote{Each value of $\eta$ has two possible choices for $b$. The choice is determined by the sign of $m$.}. Strictly speaking this quantization involves a family of orbits (associated to the invariant torus) rather than just one, although this could plausibly be avoided by a reformulation in terms of specifically the periodic orbits along the lines of \cite{berry1976closed}. Moreover, it is in some sense overconstraining, only one undetermined quantum number $n$ is still available but there are two transverse directions. One of the transverse directions from our subsequent quantization on the Penrose limit spacetime, the stable one, would thus only act to change the background null geodesic up to a zero-point contribution, which we have not yet included in \eqref{eqn:EBK}. Therefore, the background geodesic essentially determines the real part of the quasinormal mode. The other, unstable transverse direction will be associated to the radial quantum number $n$ and fixes the decay, or the imaginary part of the quasinormal mode.  \\

\begin{figure}[t!]
	\begin{center}
			\includegraphics[width=0.48\textwidth]{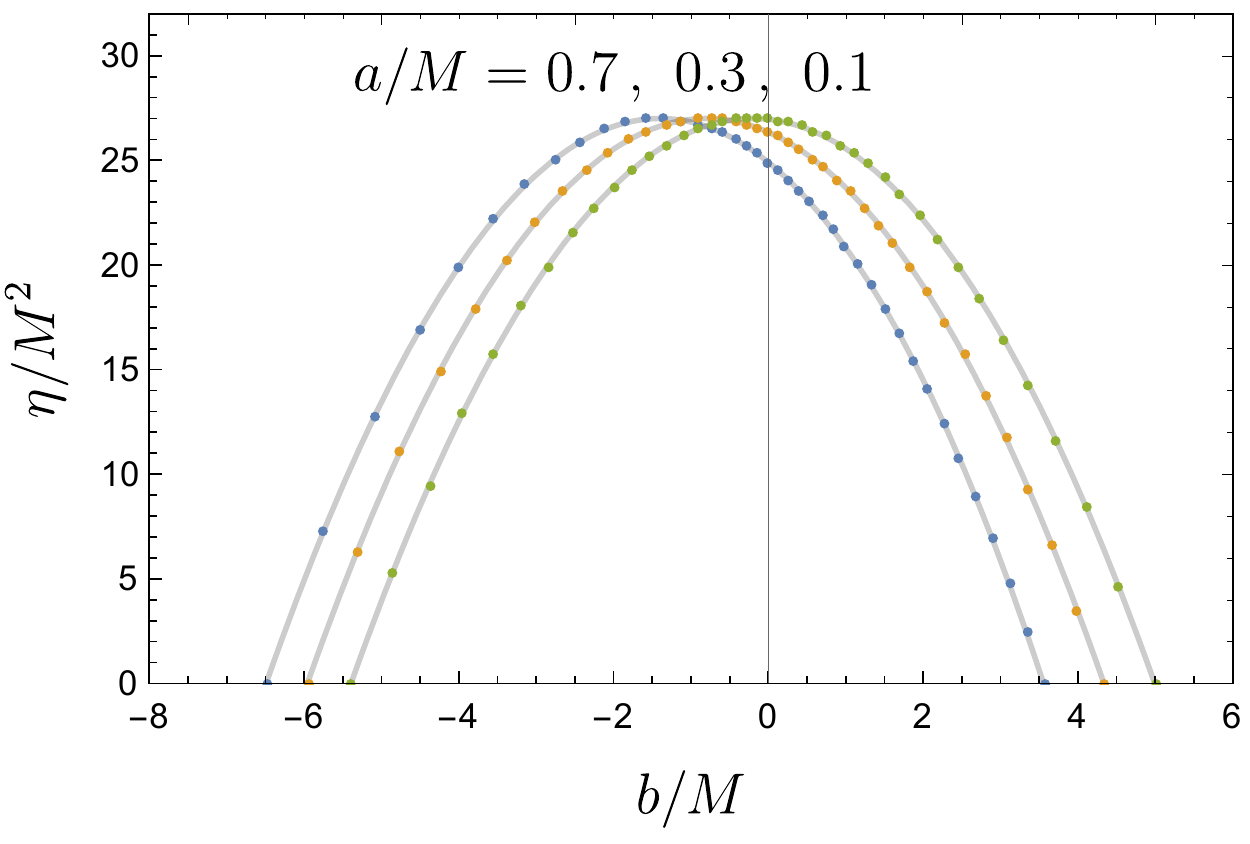} 
			\includegraphics[width=0.48\textwidth]{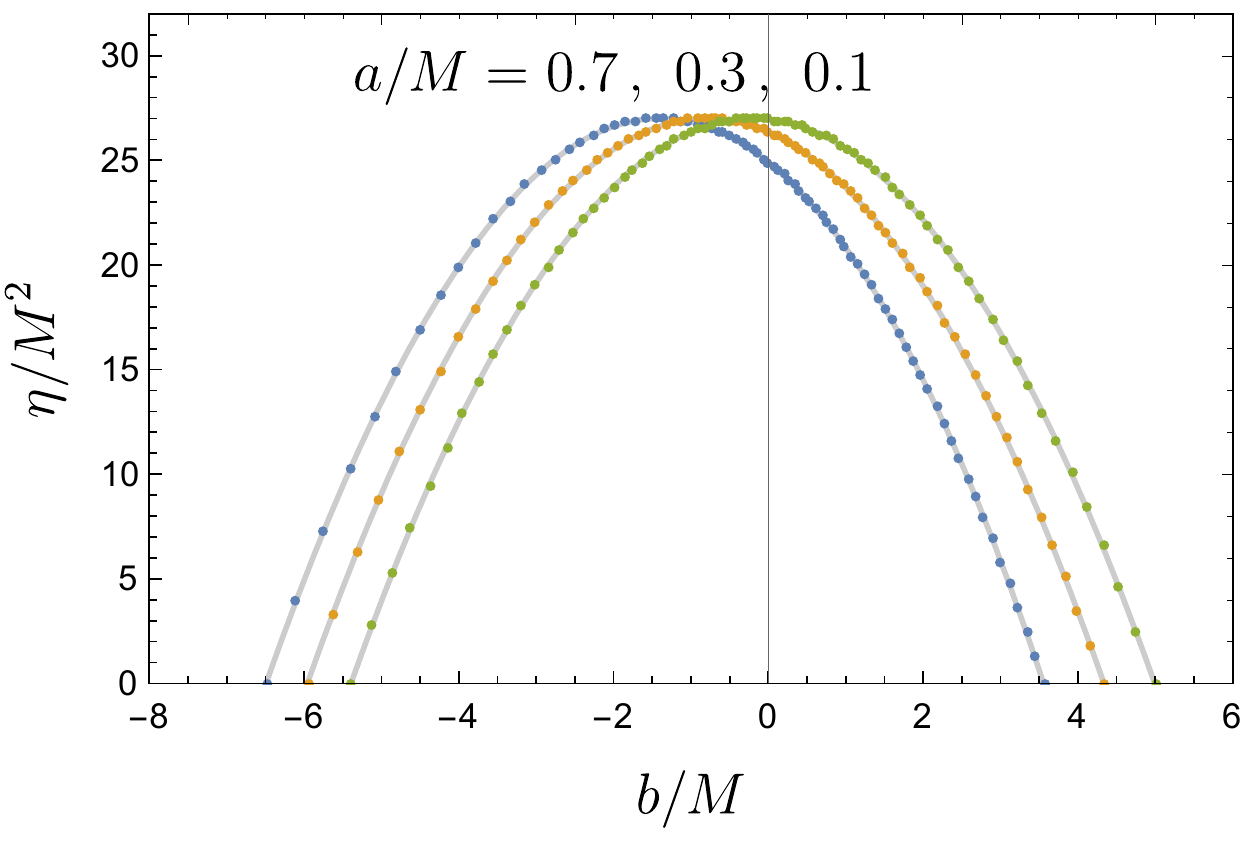}
		\caption{$b$ and $\eta$ for spherical null geodesics associated to particular mode numbers $l$, $m$ based on the (semi-classical quantization) condition \eqref{eqn:EBK} for $l=20$ (left) and $l=40$ (right) and various values of the spin ($a=0.7$, $a=0.4$ and $0.1$ left to right). The lines represent the continuous values of spherical null geodesics as approached in the geometrical optics limit. $\eta$ reaches a maximum of $27M$ at $b = -2a$ corresponding to $r = 3M$ \cite{Compere:2021bkk}.}
		\label{fig:orbitsquantumnumbers}
	\end{center}
\end{figure} 

Instead of tackling the wave equation, as in the previous section, let us start by considering the transversal part of the geodesic equation \eqref{eqn:planewavegeodesic} on \eqref{eqn:ppkerrgeneral}, that is, the classical harmonic oscillator with time-dependent frequency matrix
\begin{equation}\label{eqn:kerrplanewavegeodesic}
\ddot{x}^{i}(u) =  A_{i j}\left(u\right) x^{j}(u) \, . 
\end{equation}
This controls all aspects of the plane wave spacetime and is specifically sufficient to determine the quasinormal mode frequencies. The system \eqref{eqn:kerrplanewavegeodesic} is a pair of linear second order ordinary differential equation with periodic coefficients. Therefore, it is subject to the Floquet theorem which states that, denoting $U$ as the period and including the first derivatives in the variables $X^a = \left\lbrace x^1, \dot{x}^1,x^2,\dot{x}^2 \right\rbrace$, the solution can be written as \cite{floquet1883equations}
\begin{equation}\label{eqn:kerrfloquet}
	X^{a}(u) = M^a{}_{b}(u) X^{b}_{\flat}(u) \, , \qquad M(u) = e^{u K}  \, , \qquad \frac{d K}{du} = 0 \, , \qquad  X^{j}_{\flat}(u+U) = X^{j}_{\flat}(u) \, .
\end{equation}
In words, it can be decomposed in terms of a secular evolution fixed by the monodromy matrix $M(U)$, or the Floquet-Lyapunov exponent $K$, and a periodic piece $X^{j}_{\flat}(u)$. The Floquet-Lyapunov exponent $K$ will play the same role as the (square root) frequency matrix in the constant coefficient case and its eigenvalues will fix the the precession frequency and Lyapunov exponent of the quasinormal modes. That is, \eqref{eqn:kerrEqQNM} still holds, only now the frequency of the background orbit as well as the eigenvalues of the monodromy matrix are more difficult to determine. In general, it might not be possible to find analytic expressions for the solutions to \eqref{eqn:kerrplanewavegeodesic}. However, given the fruitful application of analytic approaches to even (the separated ordinary different equations associated to) the wave equation, for instance in terms isomonodromic deformations \cite{ohyama2006coalescent}, conformal blocks \cite{Litvinov:2013sxa,Piatek:2017fyn, Bonelli:2021uvf,Lisovyy:2021bkm, Lisovyy:2022flm, Bonelli:2022ten} or Seiberg-Witten theory \cite{Aminov:2020yma, Bianchi:2021xpr, Bianchi:2021mft} one should expect that \eqref{eqn:kerrplanewavegeodesic} is amenable to such approaches too. To make a connection to these works, it is useful to consider \eqref{eqn:kerrplanewavegeodesic} in terms of $\left\lbrace \theta, p_{\theta}\right\rbrace$ instead of $u$ where, from \eqref{eqn:kerractions} and \eqref{eqn:kerrpolar}
\begin{equation}\label{eqn:thetaelliptic}
	p_{\theta}^2 = \eta - \cos^2 \theta \left(\frac{b^2}{\sin^2 \theta}-a^2 \right) \,  .
\end{equation}
Introduce the change of variables $z = \cos{\theta}$ to write \eqref{eqn:thetaelliptic} as
\begin{equation}
\begin{aligned}
p_{z}^2 &= \frac{\eta +\left(a^2-b^2-\eta \right)z^2 -a^2 z^4 }{(1-z^2)^2} = \frac{r_0^3\left(4 a^2 M - r_0 \left(r_0-3M\right)^2\right)-2r_0\left(r_0^3-3M^2r_0+2a^2M\right)a^2 z^2-a^4 z^4}{a^2(1-z^2)^2(r_0-M)^2} \,  ,  \\ p_z^2 &= \frac{p_\theta^2}{1-z^2} \, . 
\end{aligned}
\end{equation}
This is directly related to the Seiberg-Witten curve and differential that would be associated to the separated angular Teukolsky equation \cite{Bianchi:2021xpr}. However, rather than requiring its periods for the quantization, as might be the case for the separated wave equation \cite{Aminov:2020yma}, one seems to need a monodromy of the flat connection on the curve associated to the differential equation 
\begin{equation}\label{eqn:kerrflatconnectioncurve}
\begin{aligned}
p_z \frac{d x_1}{d z} &= \left(r_0^2+a^2 z^2\right) v_1 \, ,\\
p_z  \frac{d v_1}{d z} &=\frac{12 M r^3_0 \Delta}{(r_0-M)^2 \left(r_0^2+a^2 z^2\right)^4}\left\lbrack 5\left(r_0^2-a^2 z^2\right)^2-4r_0^4\right\rbrack x_1 +  a z  \frac{12 M r^2_0 \Delta}{(r_0-M)^2 \left(r_0^2+a^2 z^2\right)^4} \left\lbrack 5\left(r_0^2-a^2 z^2 \right)^2-4 a^4 z^4  \right\rbrack  x_2 \, , \\
p_z  \frac{d x_2}{d z} &= \left(r_0^2+a^2 z^2\right) v_2 \, , \\
p_z \frac{d v_2}{d z} &=   a z \frac{12 M r^2_0 \Delta}{(r_0-M)^2 \left(r_0^2+a^2 z^2\right)^4} \left\lbrack 5\left(r_0^2-a^2 z^2 \right)^2-4 a^4 z^4  \right\rbrack   x_1 - \frac{12 M r^3_0 \Delta}{(r_0-M)^2 \left(r_0^2+a^2 z^2\right)^4}\left\lbrack 5\left(r_0^2-a^2 z^2\right)^2-4r_0^4\right\rbrack x_2  \, .
\end{aligned}
\end{equation}
In principle at least, one should be able to apply similar isomonodromy, CFT and Seiberg-Witten theory methods to \eqref{eqn:kerrflatconnectioncurve} \cite{Korotkin:1995yi, DelMonte:2022nem}. However, if no additional simplifications are found, it would be significantly more complicated. Nevertheless, contrary to first separating the wave equation and subsequently applying these methods to both angular and radial equations in isolation as, the formulation in terms of a Penrose limit \eqref{eqn:kerrflatconnectioncurve} seamlessly integrates the geometrical optics (or semi-classical) approximation of the radial problem into the residual angular problem. If a physical interpretation is to be found between the different approaches to the quasinormal modes, it seems more promising to do so in a way that self-consistently captures the entire problem, as achieved here, rather than by matching two ordinary differential equations of a special type individually. We will not pursue this further here. Instead in the following, we will perform a preliminary numerical analysis of \eqref{eqn:kerrfloquet} and compare to numerically computed quasinormal modes. \\

\begin{figure}[t!]
	\begin{center}
			\includegraphics[width=0.48\textwidth]{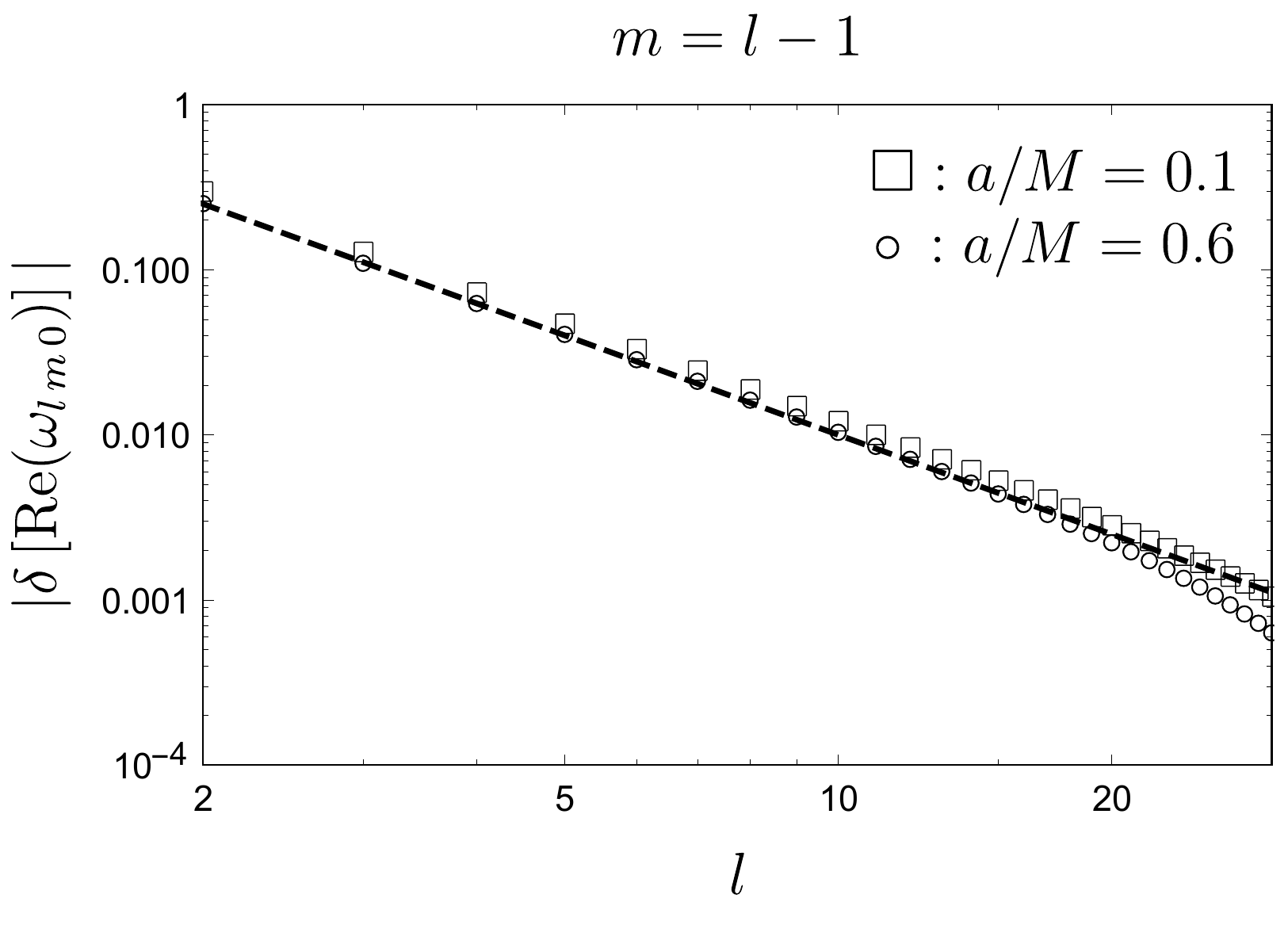} 
			\includegraphics[width=0.48\textwidth]{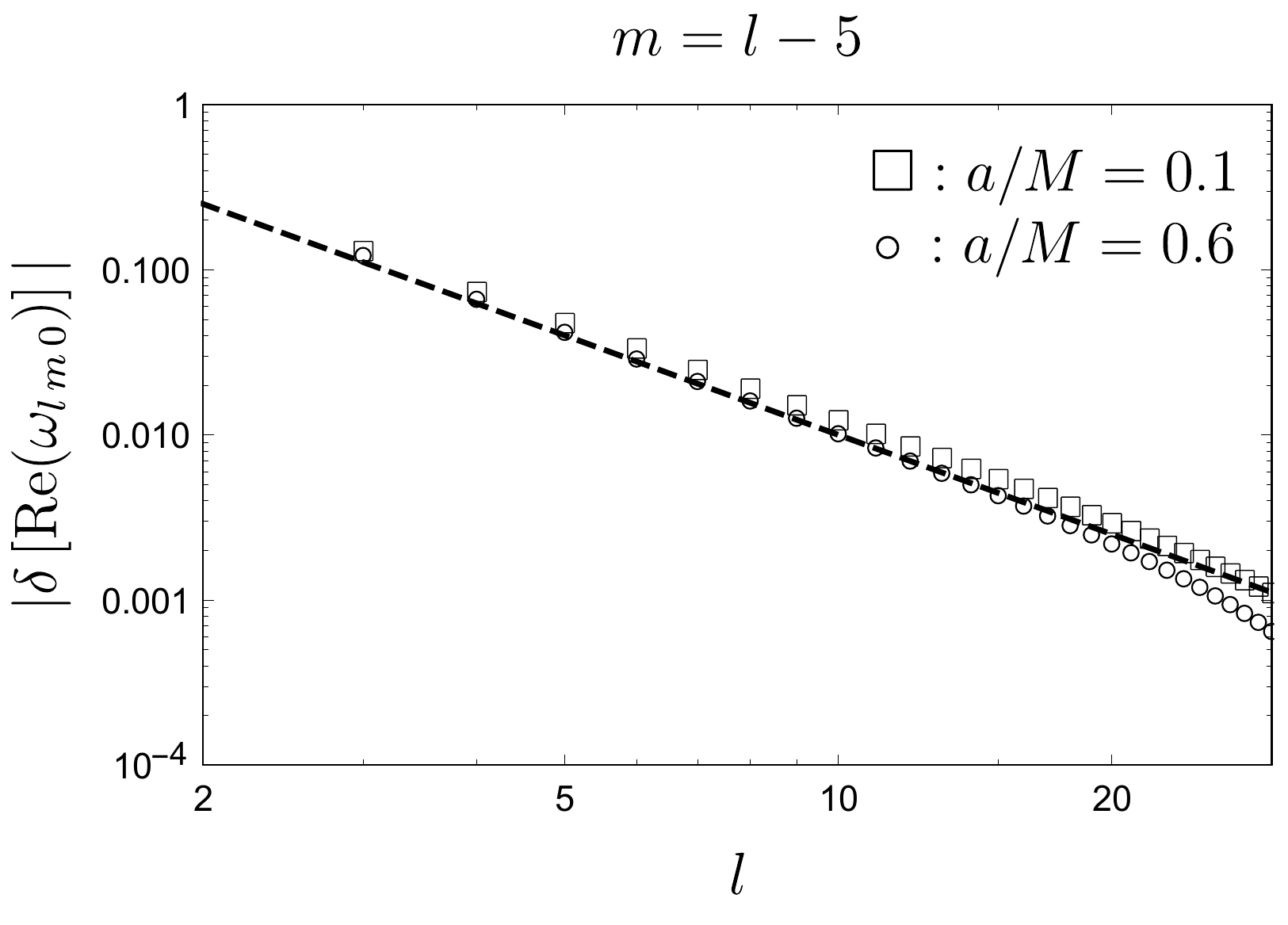} 
			\\
			\vspace{30pt}
			\includegraphics[width=0.48\textwidth]{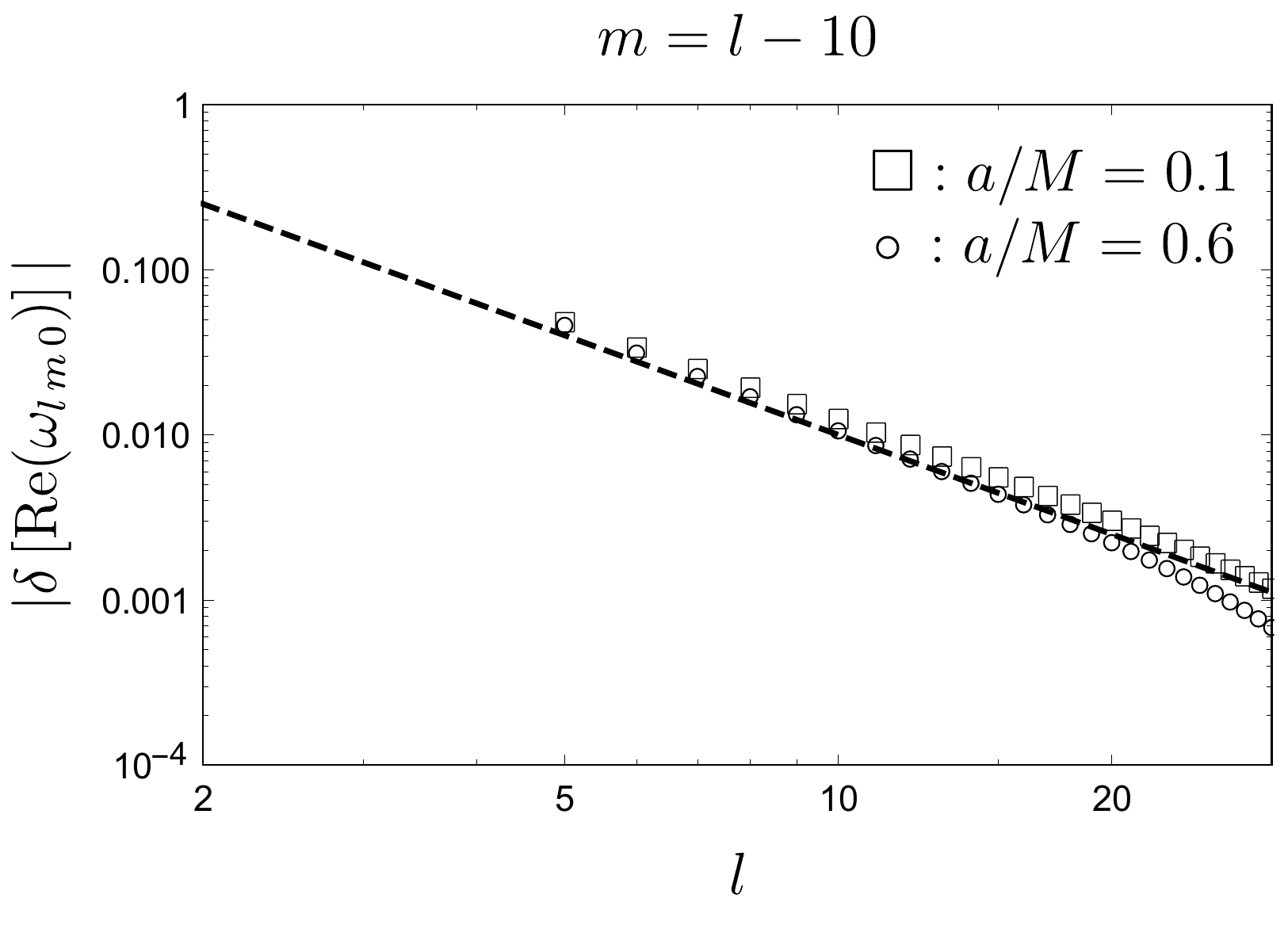} 
			\includegraphics[width=0.48\textwidth]{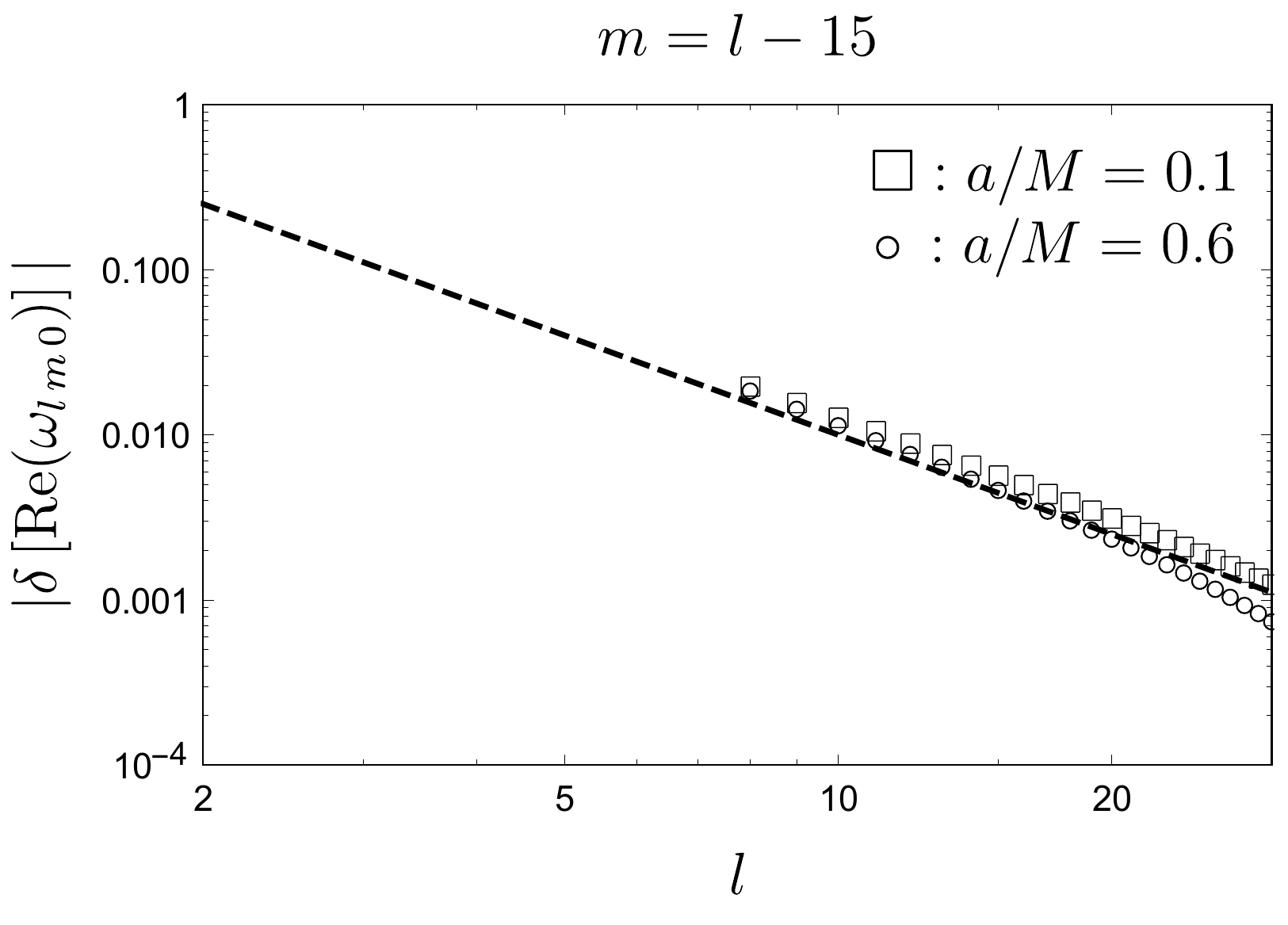} 
		\caption{Comparison of the geometrical optics approximation \eqref{eqn:kerrEqQNM} to the (real part of the) Kerr quasinormal modes with their numerical values, for the fundamental $n=0$ mode as a function of the angular mode number $l$, for different values of spin as well as azimuthal mode number $m$. Specifically, what is shown is the relative difference of the real part of the quasinormal mode between the full numerical quasinormal mode frequency and the geometrical optics approximation, which we denote as  $\delta x = \left(x_{\rm approx}-x_{\rm num}\right)/x_{\rm num}$. In \eqref{eqn:kerrEqQNM}, the leading real part of the quasinormal mode frequency $\omega_{\rm orb}$ is found from the EBK quantization condition \eqref{eqn:EBK}, while the first subleading correction $\omega_{\rm prec}$ as well as the leading imaginary part $\lambda_L$ are found from the (numerically computed) Floquet-Lyapunov exponent matrix \eqref{eqn:kerrfloquet}. The convergence with the angular mode number $l$ follows (at least) the expected $l^{-2}$-scaling as indicated by the (black) dashed reference line.}
		\label{fig:kerrapproxre}
	\end{center}
\end{figure}

\begin{figure}[t!]
	\begin{center}
		\includegraphics[width=0.48\textwidth]{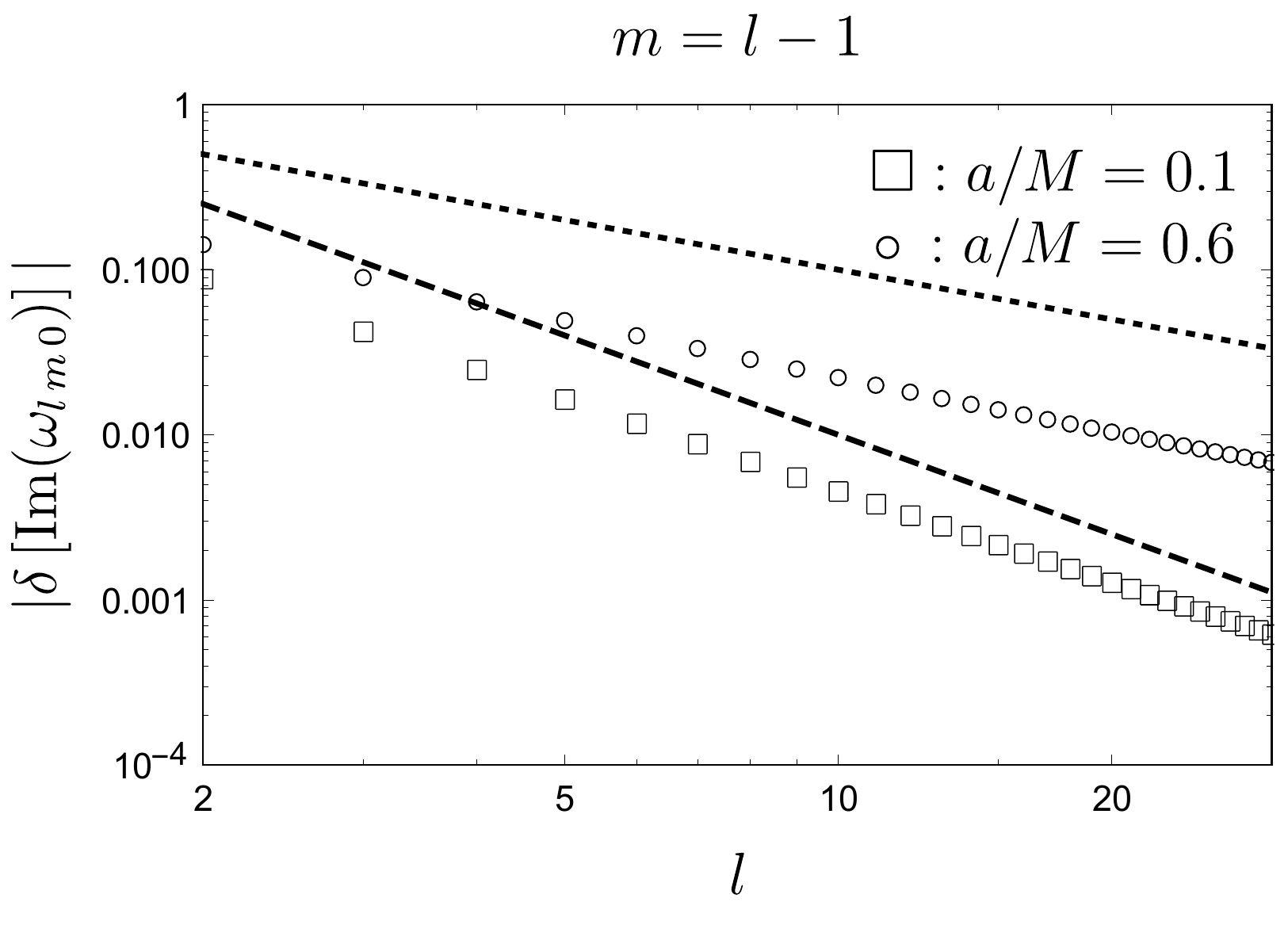} 
		\includegraphics[width=0.48\textwidth]{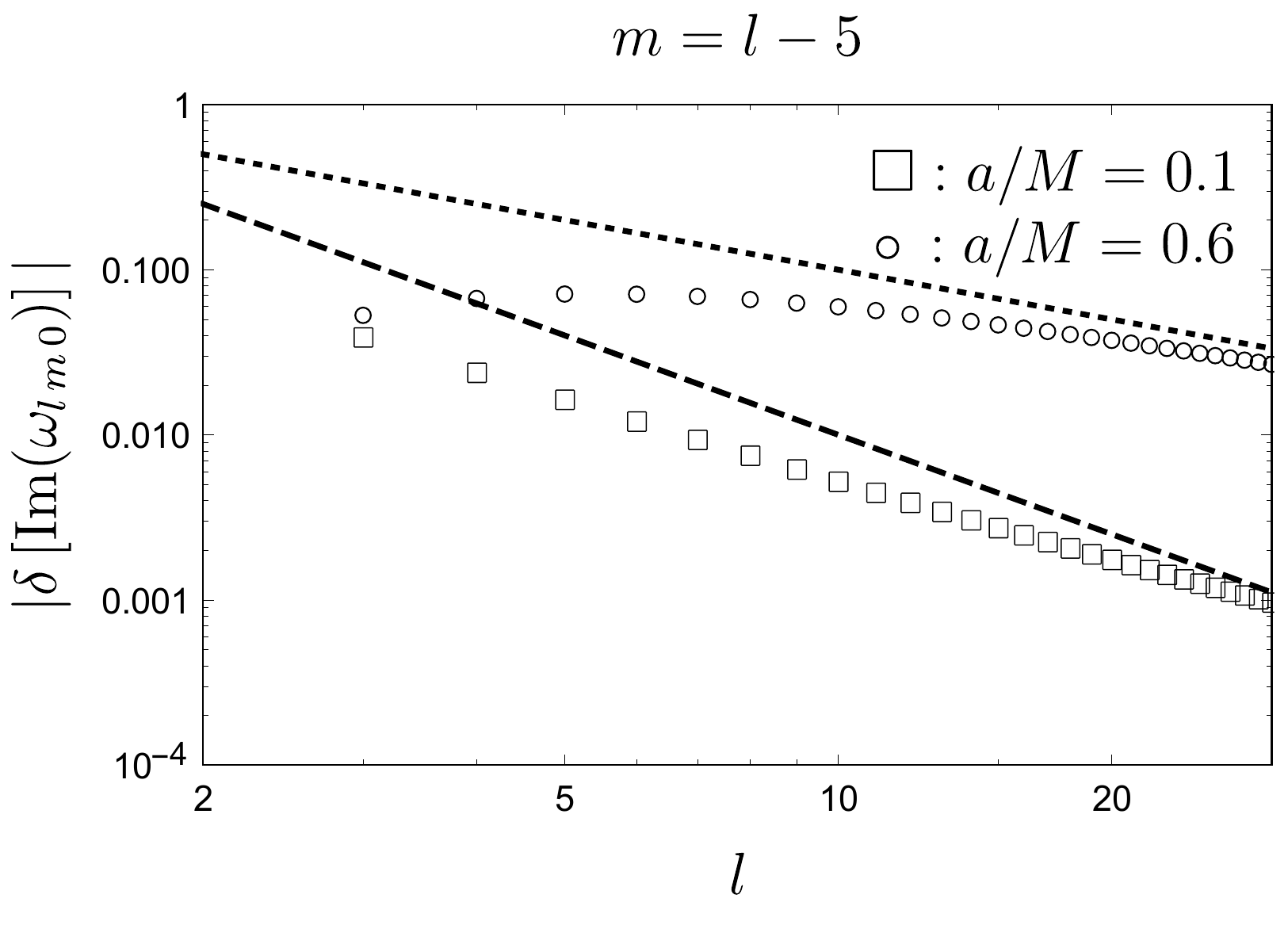} 
		\\
		\vspace{30pt}
		\includegraphics[width=0.48\textwidth]{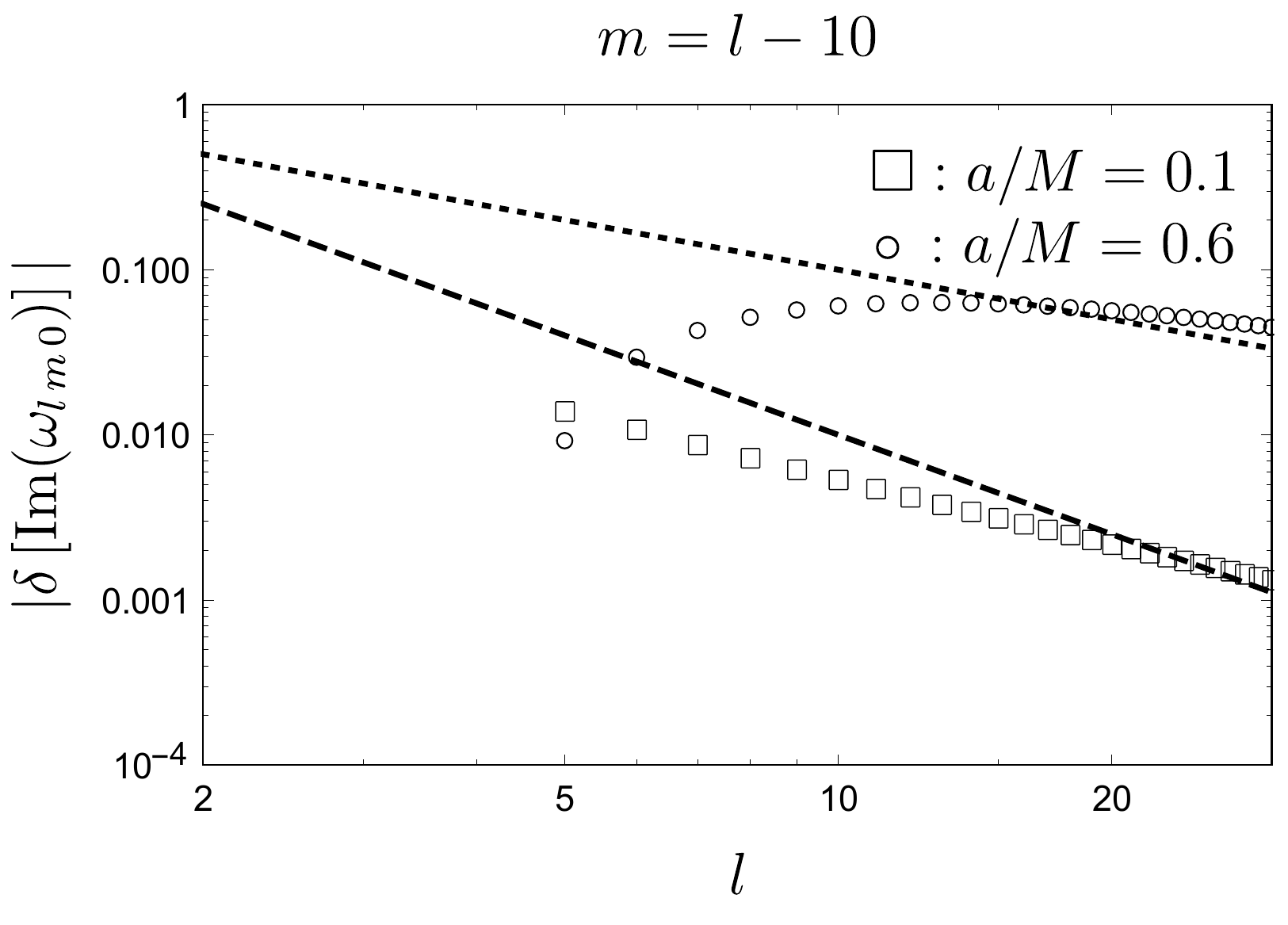} 
		\includegraphics[width=0.48\textwidth]{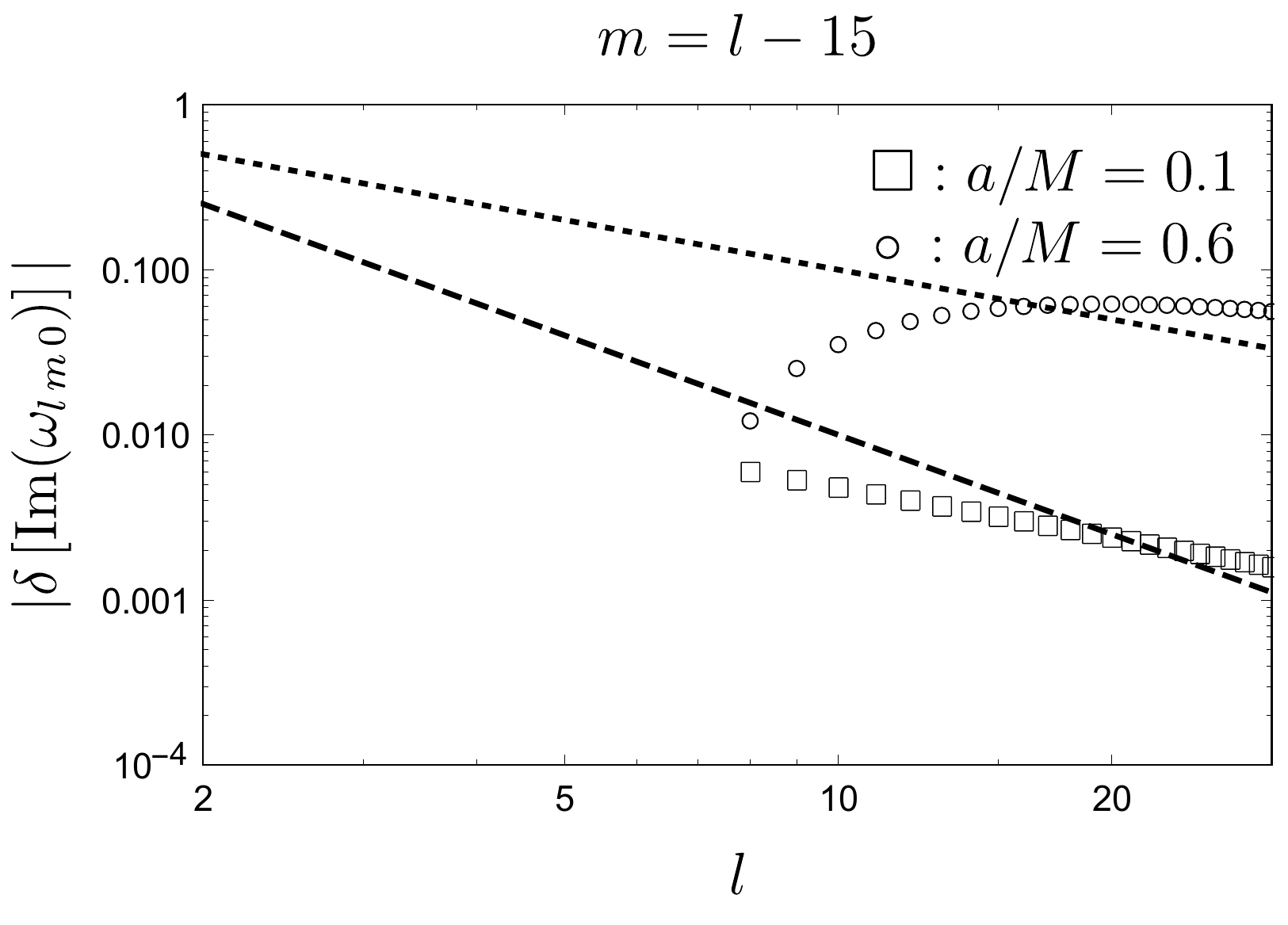} 
		\caption{Comparison of the geometrical optics approximation \eqref{eqn:kerrEqQNM} to the (imaginary part of the) Kerr quasinormal modes with their numerical values, for the fundamental $n=0$ mode as a function of the angular mode number $l$, for different values of spin as well as azimuthal mode number $m$. Specifically, what is shown is the relative difference of the imaginary part of the quasinormal mode between the full numerical quasinormal mode frequency and the geometrical optics approximation, which we denote as  $\delta x = \left(x_{\rm approx}-x_{\rm num}\right)/x_{\rm num}$. In \eqref{eqn:kerrEqQNM}, the leading real part of the quasinormal mode frequency $\omega_{\rm orb}$ is found from the EBK quantization condition \eqref{eqn:EBK}, while the first subleading correction $\omega_{\rm prec}$ as well as the leading imaginary part $\lambda_L$ are found from the (numerically computed) Floquet-Lyapunov exponent matrix \eqref{eqn:kerrfloquet}. The convergence with the angular mode number $l$ follows (at least) the expected $l^{-1}$-scaling as indicated by the (black) dotted reference line. However, as was already observed for the Schwarzschild black hole in Figure \ref{fig:Schwarzschildqnms}, the convergence can be better at low spin, even approaching the $l^{-2}$ for the equatorial $l \sim m$ modes as indicated by the (black) dashed reference line.}
		\label{fig:kerrapproxim}
	\end{center}
\end{figure}

\begin{figure}[t!]
	\begin{center}
		\includegraphics[width=0.48\textwidth]{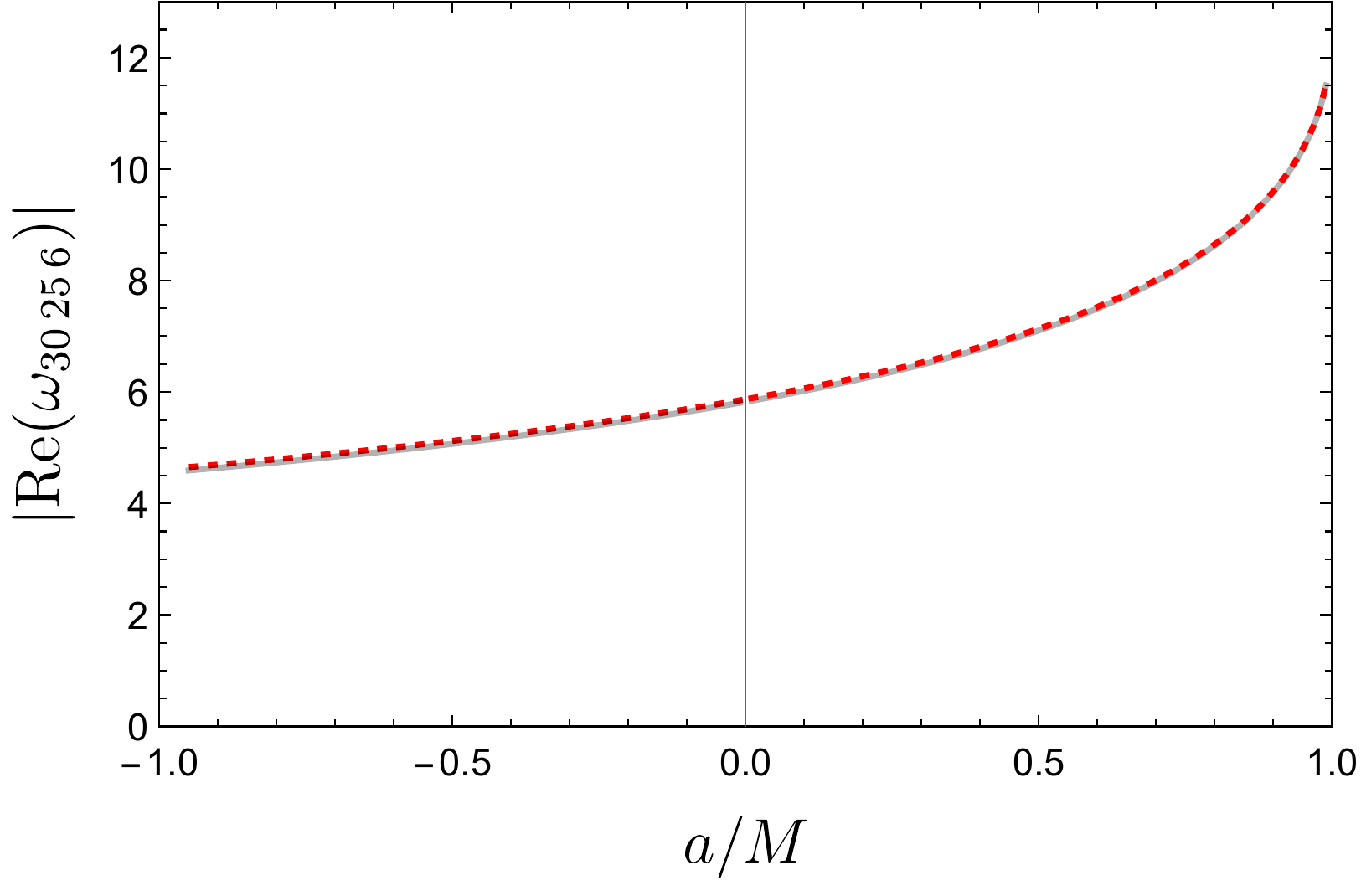} 
		\includegraphics[width=0.48\textwidth]{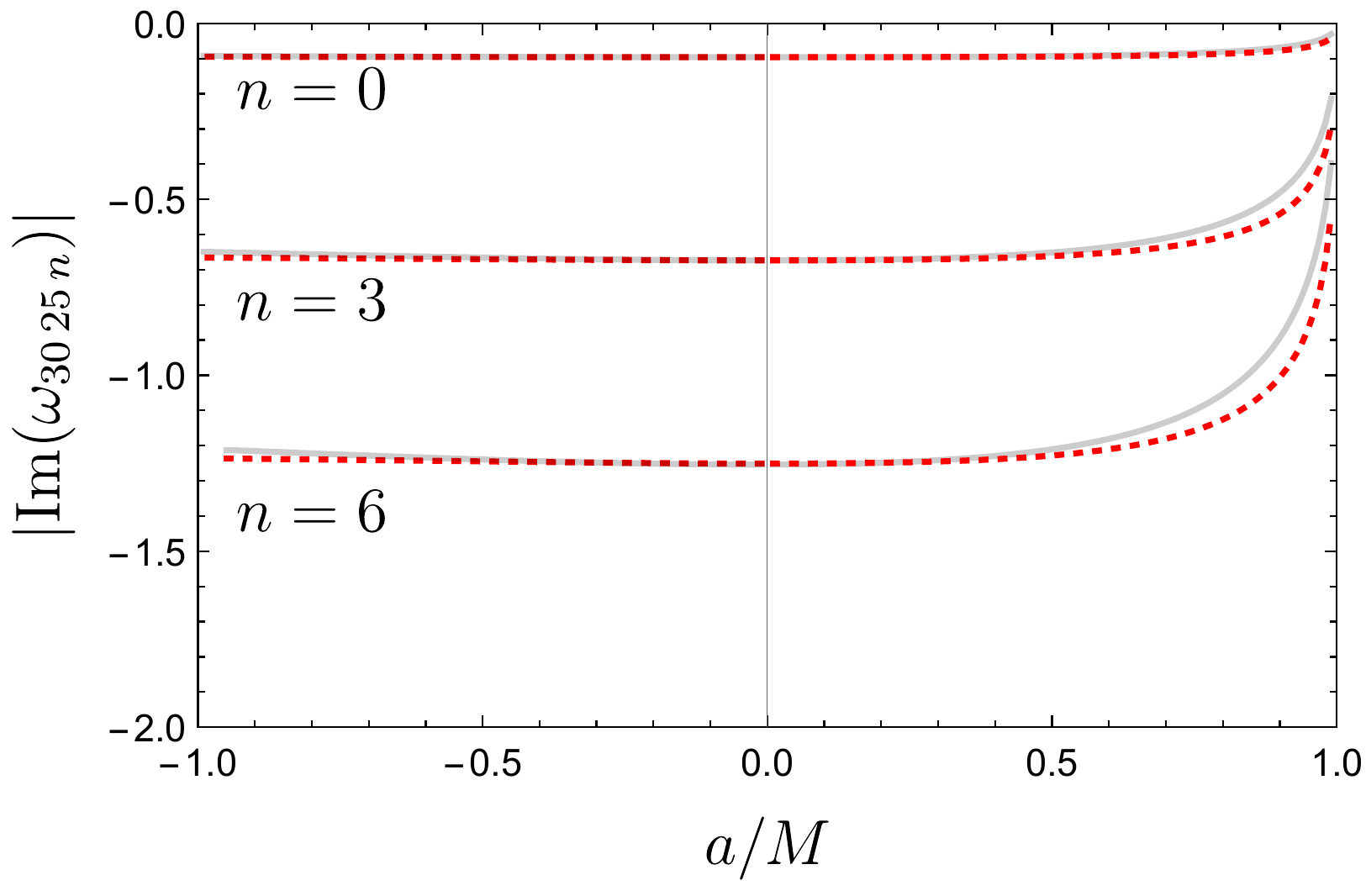}
		\caption{Comparison of the geometrical optics approximation \eqref{eqn:kerrEqQNM} (red, dashed) to the numerically computed Kerr quasinormal mode frequencies (gray, full) for $l=30$, $m=25$ as a function of the black hole angular momentum at different values of the overtone number $n$ (for the imaginary part where the dependence is non-negligible). In \eqref{eqn:kerrEqQNM}, the leading real part of the quasinormal mode frequency $\omega_{\rm orb}$ is found from the EBK quantization condition \eqref{eqn:EBK}, while the first subleading correction $\omega_{\rm prec}$ as well as the leading imaginary part $\lambda_L$ are found from the (numerically computed) Floquet-Lyapunov exponent matrix \eqref{eqn:kerrfloquet}. The geometrical optics approximation in the corotating extremal limit correctly captures the behavior of the zero-damped quasinormal modes, as observed in \cite{Yang:2012he}. The relative differences of this example are shown in Figure \ref{fig:kerrspindiff}. The convergence with $l$ at different values of $l-m$ is shown in Figures \ref{fig:kerrapproxre}-\ref{fig:kerrapproxim}.}
		\label{fig:kerrspin}
	\end{center}
\end{figure}

\begin{figure}[t!]
	\begin{center}
		\includegraphics[width=0.48\textwidth]{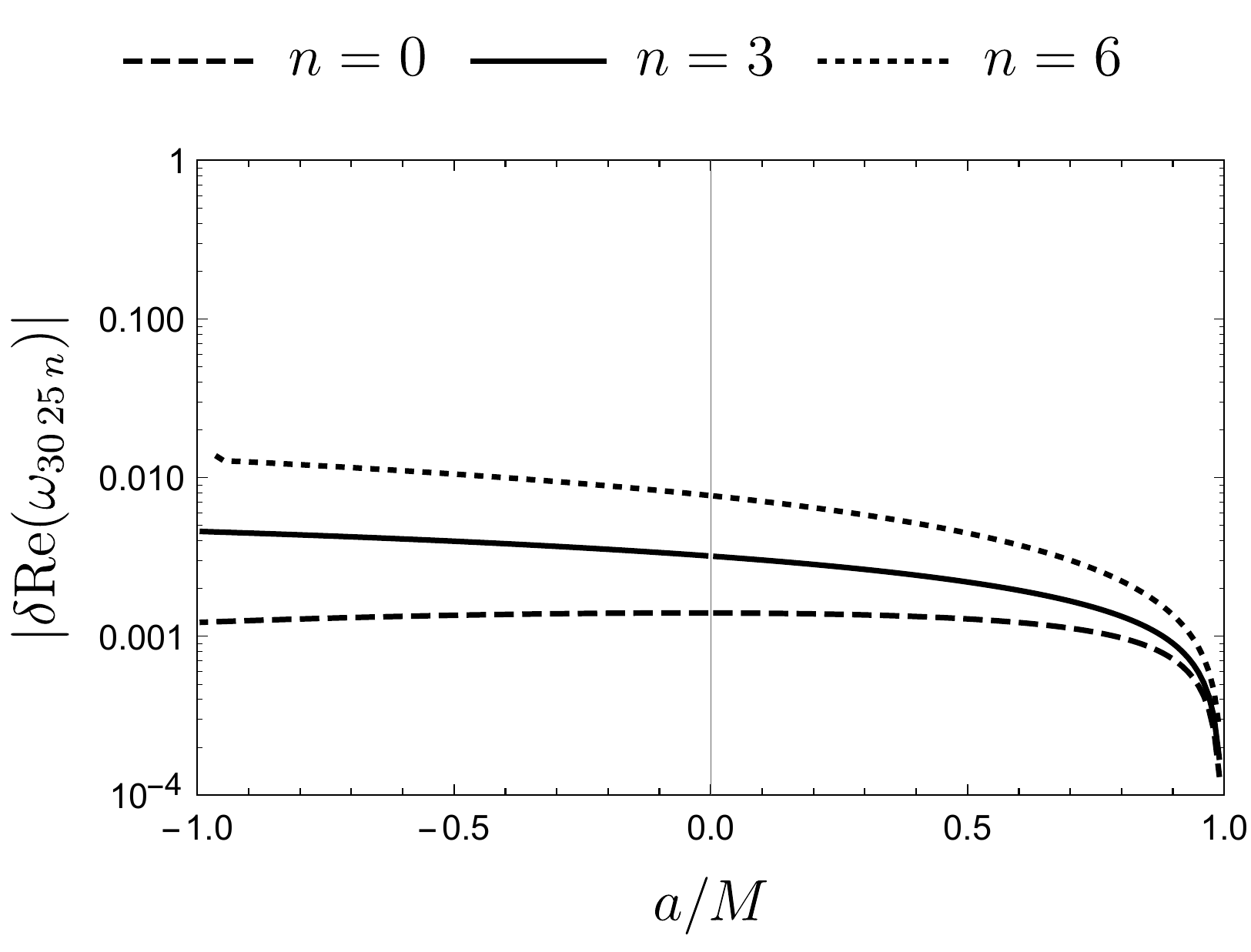} 
		\includegraphics[width=0.48\textwidth]{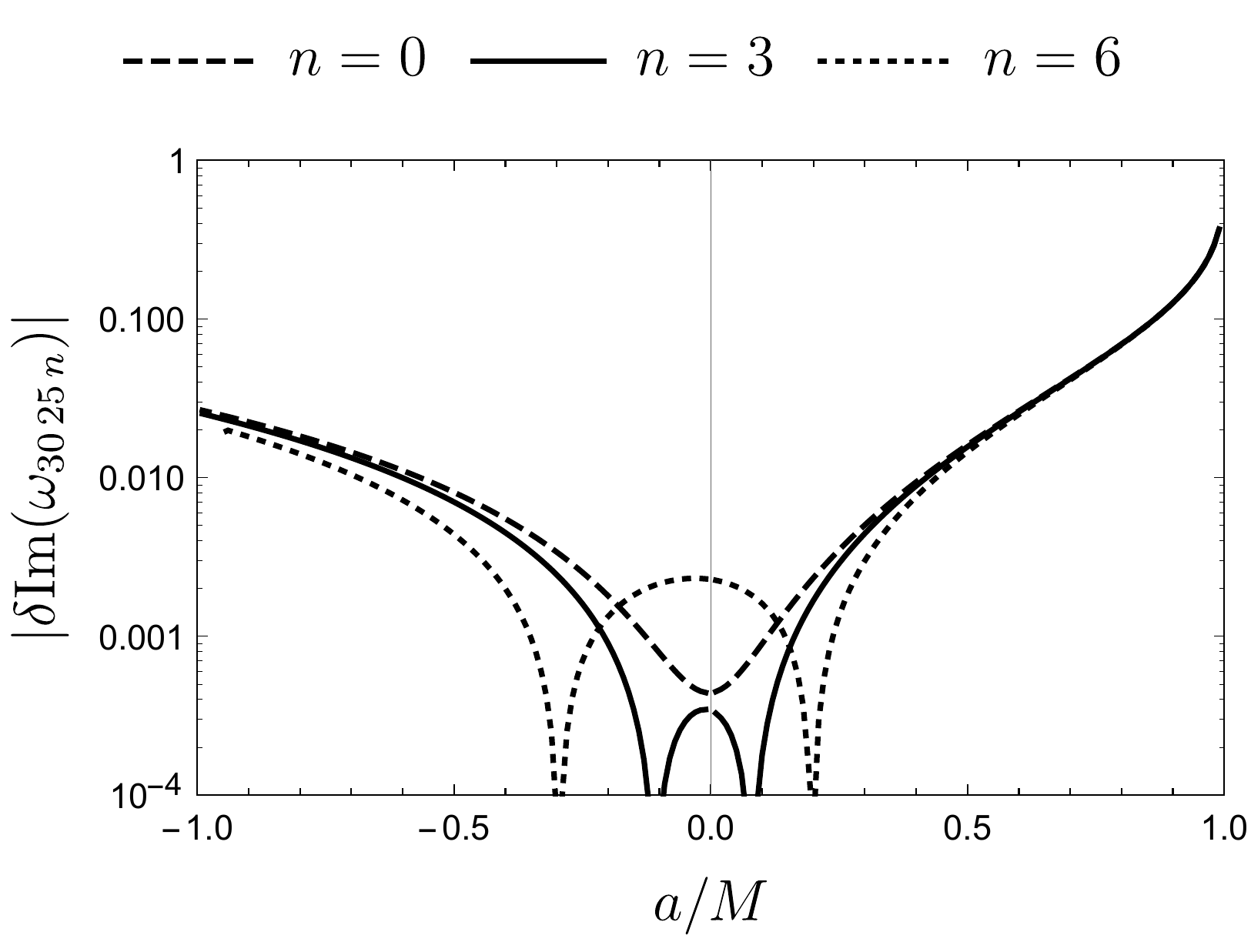}
		\caption{Comparison of the geometrical optics approximation \eqref{eqn:kerrEqQNM} to numerically computed Kerr quasinormal mode frequencies for $l=30$ and $m=25$ as a function of the (dimensionless) black hole angular momentum. For the real part of the quasinormal mode frequency (left), the dependence on the overtone $n$ is small and we show only $n=6$. For the imaginary part (right), the dependence on this overtone number is more significant and several values ($n=0, 3, 6$ are shown. For a general quantity $x$, we denote such a relative difference as $\delta x = \left(x_{\rm approx}-x_{\rm num}\right)/x_{\rm num}$. In \eqref{eqn:kerrEqQNM}, the leading real part of the quasinormal mode frequency $\omega_{\rm orb}$ is found from the EBK quantization condition \eqref{eqn:EBK}, while the first subleading correction $\omega_{\rm prec}$ as well as the leading imaginary part $\lambda_L$ are found from the (numerically computed) Floquet-Lyapunov exponent matrix \eqref{eqn:kerrfloquet}. The geometrical optics approximation in the corotating extremal limit correctly captures the behavior of the zero-damped quasinormal modes, as observed in \cite{Yang:2012he}. For the imaginary part, which goes to zero in that limit, this is more clearly seen in Figure \ref{fig:kerrspin}. The convergence with $l$ at different values of $l-m$ is shown in Figures \ref{fig:kerrapproxre}-\ref{fig:kerrapproxim}.}
		\label{fig:kerrspindiff}
	\end{center}
\end{figure}

To numerically extract $K$, we first use the analytic expressions for the geodesic evolution of $\theta$ as given in \cite{Gralla:2019ceu} to find the explicit form of $A_{ij}$, in terms of (Mino) time. In the language of \cite{Gralla:2019ceu}, we are interested in ordinary polar motion which is symmetric across the equator (as opposed to vortical motion). For this case, in terms of the Mino time $\tau$,
\begin{equation}\label{eqn:minotime}
\frac{d u}{d \tau} = \Sigma \, ,
\end{equation}
 one has \cite{Gralla:2019ceu}
\begin{equation}
\begin{aligned}
	\frac{\cos{\theta}}{|\cos{\theta_{+}|}} &= -\nu_{\theta} {\rm sn}\left(\sqrt{-\cos^2{\theta_{-}} a^2}\left(\tau + \nu_{\theta} \cG^s_\theta \right) \left| \frac{\cos^2\theta_{+}}{\cos^2{\theta_-}} \right. \right) \, , \\ \cG^s_\theta = \int_{\theta_0}^{\theta}  \frac{d \theta'}{\sqrt{\Theta(\theta')}} &= -\frac{1}{\sqrt{-\cos^2{\theta_{-}} a^2}} {\rm Fe} \left(\arcsin\left(\frac{\cos{\theta}}{|\cos{\theta_{+}|}}\right)\left| \frac{\cos^2\theta_{+}}{\cos^2{\theta_-}} \right. \right) \, ,
	\end{aligned}
\end{equation}
with ${\rm sn}$ the Jacobi elliptic sine function, ${\rm Fe}$ an incomplete elliptic integral, $\nu_{\theta}$ a sign depending on the direction of the motion and the ``slashed integral'' is defined such that it is monotonically increasing along the trajectory. We simultaneously rescale $x_1$ and $x_2$ to find an equation of the same form as \eqref{eqn:kerrplanewavegeodesic}
\begin{equation}\label{eqn:kerrplanewavegeodesicmino}
	\frac{d^2y^{i}}{d \tau^2}(\tau) =  B^{i}_{j}(\tau)  y^{j}(\tau) \, , \qquad x^i = \sqrt{\Sigma(\tau)} y^i \, .
\end{equation}
The change of variables from $x$ to $y$ does not change the monodromy due to the periodicity of $\Sigma$. By numerically extracting the monodromy over one period $T_{\tau}$ of Mino time\footnote{For the numerical integration, we use an explicit Runga-Kutta routine.  However, given the prevalence of similar Floquet problems in engineering this can no doubt be improved upon \cite{lust2001improved}.}, and specifically the eigenvalues $\pm \sqrt{\pm k_{\pm}}$ of the (Floquet-Lyapunov) exponent matrix $K_{\tau}$ (although with respect to Mino time), we then finally find the precession frequency and Lyapunov exponent in Boyer-Lindquist time as
\begin{equation}
\omega_{\rm prec} = \sqrt{-k_-}\frac{T_{\tau}}{T_t} \, , \qquad \lambda_L = \sqrt{k_+}\frac{T_{\tau}}{T_t} \, , 
\end{equation}
with the Mino time period $T_{\tau}$ and ordinary time period $T_t$ given by \cite{Gralla:2019ceu}
\begin{equation}
	T_{\tau} = \frac{4}{\sqrt{-\cos^2{\theta_{-}} a^2}} {\rm Ke} \left( \frac{\cos^2\theta_{+}}{\cos^2{\theta_-}} \right) \, , \qquad T_{t} = r^2 \frac{r+3M}{r-M} T_{\tau} -\frac{8 a^2 \cos^2{\theta_{+}}}{\sqrt{-\cos^2{\theta_{-}} a^2}} {\rm Ee}'\left( \frac{\cos^2\theta_{+}}{\cos^2{\theta_-}} \right) \, .
\end{equation}
As a consistency check, we have verified that the correct equatorial limit is recovered. In addition, we find the expected scaling of the approximation with $l$ in the comparison with numerically computed quasinormal modes, as shown in Figure \ref{fig:kerrapproxre} and Figure \ref{fig:kerrapproxim}. This is to say, the relative error to the real part of the quasinormal mode frequency decreases as $\mathcal{O}\left(l^{-2}\right)$ while that to the imaginary part decreases as $\mathcal{O}\left(l^{-1}\right)$. The difference results from the fact that the leading imaginary part occurs only at the first subleading order of the geometrical optics approximation. In Figure \ref{fig:kerrapproxre}, one can observe that the improvements in accuracy is in fact somewhat better for larger $a/M$, which could be a reflection of how well the real part of the extremal limit $a/M \to 1$ is captured for corotating, zero-damped modes. The similar deterioration in the imaginary part, despite this being captured as well, would then result from the rate at which the limit is approached, which as it is going to zero, can significantly impact the relative difference. We illustrate the dependence of the approximation on the overtone number $n$ and on the black hole angular momentum in Figures  \ref{fig:kerrspin} and \ref{fig:kerrspindiff} for the example $l = 30$, $m=25$. The correct behavior but the different approach to the near-extremal zero-damped regime, which could explain the features just noted for the $a/M$ dependence in Figure \ref{fig:kerrapproxre} and Figure \ref{fig:kerrapproxim}, are then best seen for this example in Figure \ref{fig:kerrspin}. In addition, we find the geometrical optics approximation generally works better for the real part at lower $n$ and higher spin. As could also be seen from the previously discussed equatorial modes $l \sim |m|$ in Figure \ref{fig:kerrequatorialapprox}. The deterioration of the relative error in the approximation with the angular momentum in the imaginary part on the other hand is most dramatic away from the equatorial modes with $|m|\sim l$. \\

 The (unapproximated) numerical quasinormal mode frequencies that were used for these comparisons are based on a combination of the results of \cite{Berti:2005ys,Berti:2009kk}, as well as the ``qnm'' \emph{python}-code by L. Stein \cite{Stein:2019mop}, which itself uses an approach by Cook-Zalutskiy \cite{Cook:2014cta,Cook:2016fge,Cook:2016ngj} and, what is known in the context of quasinormal modes as, Leaver's method \cite{nollert1993quasinormal,leaver1985analytic} to numerically solve the coupled eigenvalues problem associated to the separated Teukolsky equation.  We postpone a more complete analysis, including of how well the wavefunction is approximated, for quasinormal modes as well as for wavefunctions associated to more general null geodesics, to future work. Given the astrophysical and observational importance of this case as well as the excellent understanding of the null geodesics \cite{Gralla:2019ceu, Compere:2021bkk}, the associated parallel frames \cite{Kubiznak:2008zs} and solutions to the wave equation \cite{Pound:2021qin, Sasaki:2003xr}, we believe such a dedicated analysis is both merited and possible.

\clearpage
\newpage
 
 \section{Conclusion and outlook}\label{sec:conclusion}
 
The purpose of the geometrical optics approximation is to construct (approximate) solutions of wave equations from worldlines or ``classical'' particle dynamics. For the wave equation on a rotating black hole background in the four-dimensional theory of general relativity, the possible null geodesics contributing to such an approximation for massless wave equations, that satisfy the quasinormal mode boundary conditions, appear to be dominated by the bound null geodesics of the lightring. Single-valuedness of the resulting wave-functions then imposes the Bohr-Sommerfeld or Einstein-Brillouin-Keller quantization of the associated classical invariant tori in phase space. Such a quantization for the bound null geodesics yields the leading geometrical optics approximation to quasinormal modes. The first correction comes from the linearized dynamics around the invariant tori, or transverse harmonic oscillators associated to the geodesic deviation \cite{voros1976semi}. This underlies the structure and symmetries of quasinormal modes in the geometrical optics approximation. \\

 Here, we have used that the transverse harmonic oscillators governing the leading amplitude correction in the geometrical optics approximation to the wave dynamics on a spacetime can be geometrically realized as part of that spacetime, by the plane wave that arises in the Penrose limit. Alternatively, the Eisenhart-Duval lift of those harmonic oscillators is physically part of the geometry, bringing along the associated wavefunctions, symmetries and symplectic geometry. As a result, we provide a geometrical perspective on previous work related to black hole perturbations in the geometrical optics approximation  \cite{Yang:2012he,Hadar:2022xag,Kapec:2022dvc}. This point of view emphasizes that spacetime from a highly relativistic point of view can be thought of as locally a plane wave, an old idea that should remain useful when taking into account backreaction \cite{bonnor1969gravitational,bonnor1970spinning,Aichelburg:1970dh,penrose1976any,Dray:1984ha}. Given the observational interest to move away from the strict quasinormal mode regime towards the merger of black holes \cite{Gleiser:1998rw,Zlochower:2003yh, Pazos:2006kz, Cheung:2022rbm,Sberna:2021eui,Lagos:2022otp}, we therefore believe a promising direction of future research is to use the Penrose limit point of view taken here to investigate nonlinear effects. Similarly, it could be of interest in the study of quasinormal mode excitation  \cite{Berti:2006wq,Dorband:2006gg, Dolan:2011fh,Zhang:2013ksa,Lim:2019xrb,Thornburg:2019ukt,Oshita:2021iyn}. In both cases, it would be valuable to connect more closely and in more generality to the large literature on solutions to the wave equation on a rotating black hole background \cite{Sasaki:2003xr,Berti:2009kk,Poisson:2011nh,Pound:2021qin}. This could include using higher-spin fields and making contact to the Teukolsky equation \cite{Teukolsky:1973ha}, as opposed to the scalar wave equations used here, considering other modifications to the wave equations, as for example in modified theories of gravity or black holes \cite{Hollowood:2009qz,Mathur:2005zp, Fransen:2020prl, Giddings:2020zso,Cano:2020cao, Cano:2021myl,Li:2022pcy,Hussain:2022ins, Volkel:2022khh}, the introduction of sources \cite{Green:2019nam,Toomani:2021jlo}, investigating how to describe caustics as well as the propagator and scattering on the black hole background more generally \cite{matzner1968scattering,Dolan:2008kf,Ottewill:2009uj,Casals:2009zh,Wardell:2014kea}. To do so, it will be important to understand  the matched asymptotic expansion that embeds the plane wave into the full spacetime, and the respective interplay between the ``far-zone'' outer spacetime and the ``near-zone'' plane wave. In addition, colliding plane wave spacetimes could potentially play a role to resolve certain types of caustics or interactions. Similar investigations are conceivably useful in an electromagnetic context, for black hole imaging or lensing \cite{Perlick:2004tq, Cunha:2018acu}. \\
 
 The Penrose limit point of view as explored here in the context of black hole perturbation theory ultimately does not provide any new information. It is simply a change of perspective. Whereas the geometrical optics approximation translates a field theory problem into transport along worldlines, the Penrose limit approach to the geometrical optics approximation re-expresses these transport problems again in terms of field-theory on a simple plane wave spacetime. These plane wave spacetimes have a long history and are well-studied \cite{brinkmann1925einstein, rosen1937plane,bondi1957plane,bondi1959gravitational,kundt1962exact,Gibbons:1975kk,harte2013tails,griffiths2016colliding}. Therefore, using this translation, a large body of field theory tools going from scattering amplitudes in curved spacetime \cite{Hollowood:2011yh, dinu2012infrared, ilderton2013scattering,Adamo:2017nia, Adamo:2022rmp,Adamo:2022qci,Cheung:2022pdk} to perhaps even string theory and holography \cite{Horowitz:1990sr,Nappi:1993ie, Berenstein:2002jq,Mann:2003qp,Pankiewicz:2003pg,Plefka:2003nb,Jevicki:2005ms} could become available and be applied in the varied and broad range of physics questions amenable to the geometrical optics approximation. \\

\section*{Acknowledgements}

I gratefully acknowledge the support of the Heising-Simons Foundation grant \#2021-2819. I would also like to thank G. Comp\`ere, D. Kapec and A. Lupsasca for useful correspondence, A. Holguin and A. Sivaramakrishnan for related discussions as well as R. Mancilla for lending me an interesting book. This work makes use of the Black Hole Perturbation Toolkit \cite{BHPToolkit}.

 \newcommand{\noop}[1]{}
\providecommand{\href}[2]{#2}\begingroup\raggedright\endgroup

\bibliographystyle{utphys}

\end{document}